\documentclass{JHEP3}
\usepackage{amsmath}
\usepackage[numbers,sort&compress]{natbib}
\usepackage{graphicx}
\usepackage{url}
\usepackage{float}
\usepackage{afterpage}
\usepackage[labelfont=bf,font=small,width=.9\textwidth]{caption}
\DeclareMathOperator{\Real}{Re}
\DeclareMathOperator{\Imag}{Im}

\makeatletter
\renewcommand\@ENVwarn[1]{}
\makeatother

\title{Exploring colourful holographic superconductors}

\author{Kasper Peeters, Jonathan Powell and Marija Zamaklar\\
Department of Mathematical Sciences,\\
Durham University,\\
South Road,\\
Durham DH1 3LE,\\
United Kingdom.\\
~\\
\email{kasper.peeters@durham.ac.uk}\\
\email{jonathan.powell@durham.ac.uk}\\
\email{marija.zamaklar@durham.ac.uk}}

\abstract{We explore a class of holographic superconductors built
  using non-abelian condensates on probe branes in conformal and
  non-conformal backgrounds. These are shown to exhibit behaviour of
  the specific heat which resembles that of heavy fermion compounds in
  the superconducting phase. Instead of showing BCS-like exponential
  behaviour, the specific heat is polynomial in the temperature. It
  exhibits a jump at the critical temperature, in agreement with
  real-world superconductors. We also analyse the behaviour of the
  energy gap and the AC and DC conductivities, and find that the
  systems can be either semi-conducting or metallic just above the
  critical temperature.}

\preprint{DCPT-09/43}

\begin{document}
\section{Introduction and summary}

An intriguing new application of the correspondence between string
theory and gauge theory is the study of strongly coupled systems which
exhibit superconductivity. Apart from suggesting an alternative point
of view on non-BCS superconductors, these might also help to develop a
better understanding of quantum critical points and strongly coupled
superconductors for which so far a satisfactory theory is
lacking~\cite{Hartnoll:2008vx}. Moreover, these constructions provide
an interesting new arena in which to explore and test the string/gauge
theory correspondence. We refer the reader to
\cite{Hartnoll:2009sz,Herzog:2009xv} for more background information
and references.

In the original model of~\cite{Hartnoll:2008vx}, a holographic
superconductor was constructed by considering an Abelian Higgs model
in the background of an anti-de-Sitter black hole. A stable condensate
of the scalar field can occur because of nontrivial coupling of the
scalar to the gauge potential, in combination with nontrivial
asymptotics~\cite{Gubser:2008wv} or because of the nontrivial
near-horizon geometry which destabilises the asymptotically stable
mode~\cite{Hartnoll:2008vx}. This system exhibits an energy gap which
scales linearly with the critical temperature, and a
frequency-dependent conductivity which shows some similarity with
real-world superconductors. A full analysis involving back-reaction of
the Abelian-Higgs fields on the gravitational background has also been
carried out~\cite{Hartnoll:2008kx}.

A different way to obtain a non-trivial condensate in a holographic
setup is to consider, instead, a non-abelian gauge field coupled to
gravity. Again, because of the non-trivial asymptotics, it is possible
to form condensates of the gauge field. For some time now, it has been
known that such configurations exist even in systems at zero
temperature (i.e.~without black hole horizon), where they are related
to vector meson condensates in the dual gauge
theory~\cite{Aharony:2007uu}. An application to $p$-wave
superconductors was developed in~\cite{Gubser:2008wv}, based on
earlier work in~\cite{Gubser:2008zu}. We will call these `colourful
superconductors' (not to be confused with colour superconductors in
QCD).

Many properties of these colourful superconductors have so far not
been analysed, and it is far from clear which real-world
superconductors they resemble most. Moreover, the analysis so far has
focussed on \emph{conformal} cases, since there is an expectation that
ideas of the gauge/gravity correspondence are perhaps most useful in
the context of quantum critical phenomena, which are often
characterised by relativistic conformal symmetry.

In the present paper we thus set out to investigate a variety of models
which exhibit non-abelian condensates. We will focus mostly on D-brane
constructions, both conformal and non-conformal, including the
Sakai-Sugimoto model. These models obviously have an origin in a
full-fledged string theory, however we will use them here without
worrying about the validity of string theory approximations.  Our
attitude is pragmatic, in the sense that we are mainly interested in
seeing how the models behave, with the eventual goal of improving the
description of holographic superconductors and bringing them closer to
real-world systems. For this reason, we will e.g.~work with
Yang-Mills action, without attempt to include DBI corrections to it,
and we will also not consider gravitational back-reaction.

After an introduction to the models under consideration, we will
first focus on their thermodynamic properties, in particular the
specific heat. For all systems which we analyse, we find evidence that
in the superconducting phase, they resemble non-BCS, heavy fermion
compounds. That is, their specific heat has a polynomial behaviour as
a function of temperature, in contrast to the exponential one seen in
BCS superconductors. At a more quantitative level, we find that $c_v
\sim T^n$ where $n \sim 2.7 - 5.5$ depending on the model. We also
demonstrate the appearance of a jump in the specific heat at~$T_c$, a
property common to both strongly and weakly coupled
superconductors. Above $T_c$, our approach only yields linear scaling
for the specific heat at constant chemical potential for the D3/D7
system, while this is present also in the D4/D8 case in the DBI
approximation~\cite{Kulaxizi:2008jx}.  The specific heat at constant
density is, for all conformal cases, known to behave
non-linearly~\cite{Karch:2008fa}.

We then continue in \S\ref{s:EM} with the analysis of electromagnetic
properties. For most systems under consideration we find behaviour
above the critical temperature which resembles that of a
semi-conductor, i.e.~a decreasing resistivity with increasing
temperature. In the superconducting phase, there is a small region in
temperature in which a condensate has already formed but the
resistivity is not yet entirely zero. We find that the gap scales
linearly with the critical temperature, $\omega_g\propto T_c$, but the
proportionality constant is outside the BCS range and can also be
substantially different from the ones found so far for conformal
holographic superconductors.

In two sections (\S\ref{s:realworld_cv} and \S\ref{s:realworld_em}) we
comment on the relation of our findings to real-world superconductors.


\section{Colourful superconductors}
\label{s:colourful}
\subsection{Non-abelian condensates}

Before we analyse a number of concrete holographic superconductors,
let us first summarise the general setup and some universal aspects of
the problem.  We will consider generic diagonal metrics in which the
time-time component of the line element is of the form $-g(u)\,{\rm
  d}t^2$, where~$g(u)$ has a zero at the location of the horizon. An
example of this sort is the metric of a D$p$-brane, given by
\begin{equation}
\label{e:bgmetricU}
{\rm d}s^2_{\text{D}p} = \left(\frac{u}{L}\right)^{\frac{7-p}{2}}\bigg(
\!- f_p(u) {\rm d}t^2 + \delta_{ij} {\rm d}x^i {\rm d}x^j
\bigg) + \left(\frac{u}{L}\right)^{\frac{p-7}{2}}\bigg(
\frac{{\rm d}u^2}{f_p(u)} + u^2\,{\rm d}\Omega^2_{8-p}\bigg)\,,
\end{equation}
where~$i,j=1\ldots p$\/ and~$f_p(u) = 1 - (u_T/u)^{7-p}$. 
For our numerical analysis later in the paper it is advantageous to
introduce coordinates in which the coordinate distance between the
asymptotic boundary and the horizon is finite. We will use
$r := L^2/u$.
The D$p$-brane metric then reads
\begin{equation}
\label{e:bgmetric}
{\rm d}s^2_{\text{D}p} = \left(\frac{L}{r}\right)^{\frac{7-p}{2}}\bigg(
\!- f_p(r) {\rm d}t^2 + \delta_{ij} {\rm d}x^i {\rm d}x^j
\bigg) + \left(\frac{L}{r}\right)^{\frac{p-7}{2}}\bigg(
\frac{L^4 {\rm d}r^2}{r^4\,f_p(r)} + \frac{L^4}{r^2}\,{\rm d}\Omega^2_{8-p}\bigg)\,,
\end{equation}
with~$f_p(r) = 1 - (r/r_T)^{7-p}$. The horizon is now located
at~$r=r_T$ and the asymptotic boundary is at~$r=0$. In this coordinate system the
dilaton equals
\begin{equation}
e^{-\phi} = \left(\frac{L}{r}\right)^{(p-7)(p-3)/4}\,.
\end{equation}
In addition to these D$p$-brane systems we will also consider the
$\text{AdS}_4$ black hole background, with vanishing dilaton and
metric given by
\begin{equation}
{\rm d}s^2_{\text{AdS}_4} = \left(\frac{L}{r}\right)^2 
\left( - f_3(r) {\rm d}t^2 + \delta_{ij}{\rm d}x^i{\rm d}x^j
+  \frac{{\rm d}r^2}{f_3(r)}\right)\,.
\end{equation}

For future reference, we note here that the temperature of the D$p$-brane
backgrounds, as measured by an asymptotic observer following the orbit
of the Killing vector~$\partial/\partial t$, is given by
\begin{equation}
\label{e:HawkingT}
T = \frac{7-p}{4\pi\,L} \left(\frac{L}{r_T}\right)^{\frac{5-p}{2}}\,.
\end{equation}
For the $\text{AdS}_4$ black hole the temperature is~$T = 3/(4\pi r_T)$.

If we now add a D$q$-brane to the D$p$ brane background, the
non-abelian gauge field on the D$q$-brane can
condense~\cite{Aharony:2007uu}; a similar story holds true when we add
a Yang-Mills action to the~$\text{AdS}_4$
background~\cite{Gubser:2008zu}. We will here restrict attention to an
SU(2) gauge field. The non-abelian condensate will occur if the
chemical potential~$\mu$, corresponding to the value of the $A_0$
component on the boundary, is fixed to a large enough value. The
system actually admits condensates of two different
types~\cite{Aharony:2007uu,Gubser:2008zu}.  The one which generically
turns out to have the largest energy of the two is of the form
\begin{equation}
\label{e:altcond}
A = A_0^{(3)}\sigma^3 {\rm d}t + A(r) \big( 
\sigma^{1} {\rm d}x^1 + \sigma^2 {\rm d}x^2\big) \,.
\end{equation}
This solution breaks the SU(2) symmetry to a diagonal U(1).
It is unstable~\cite{Gubser:2008wv}, and expected to decay to a lower
energy condensate. The lower energy configuration can be written in
the form\footnote{The use of the coordinate~$x^3$ is a bit misleading,
  and should \emph{not} be understood to imply that this condensate
  only occurs in models which have at least three space dimensions.}
\begin{equation}
\label{e:lowestEcondensate}
A = A_0^{(3)}\sigma^3 {\rm d}t + A_{3}^{(1)} \sigma^{1} {\rm d}x^3\,.
\end{equation}
Such a condensate is analogous to the one studied at zero temperature
in~\cite{Aharony:2007uu}, and we will return to this analogy in more
detail in section~\ref{s:zeroTSS}. In the $2+1$ dimensional case
studied in~\cite{Gubser:2008wv}, this condensate breaks both SU(2) and
rotational invariance completely.

\begin{figure}[t]
\begin{center}
\includegraphics[width=.55\textwidth]{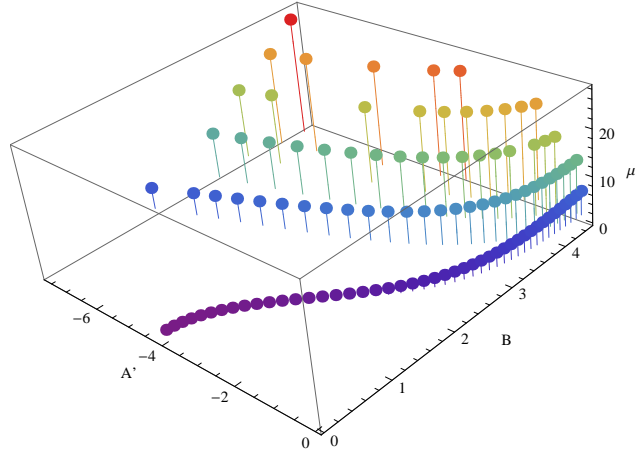}
\vspace{-2ex}
\end{center}
\caption{The structure of the condensate solution space for the D3/D7
  system, showing the chemical potential~$\mu$ as a function of the
  two parameters~$A'$ and $B$ which determine the solution at the
  horizon. Clearly visible are the various branches of
  condensates.\label{f:D3D7_branches}}
\end{figure}

We will restrict ourselves to an analysis of the Yang-Mills truncation
of the effective action on the probe brane (comments on the
limitations of this approximation will be made in
\S\ref{s:limitations}). The Lagrangian which determines the formation
of a condensate of the type~\eqref{e:lowestEcondensate} is given by
\begin{equation}
\label{e:generic_action}
L = - T_q \int_{0}^{r_T}\!{\rm d}r\, 
  \sqrt{-\hat{g}} e^{-\phi} \Big[
  \hat{g}^{00}\hat{g}^{rr} (\partial_r A_0^{(3)})^2
+ \hat{g}^{33}\hat{g}^{rr} (\partial_r A_{3}^{(1)})^2
+ 4 \hat{g}^{00} \hat{g}^{33} (A_{0}^{(3)} A_{3}^{(1)})^2
\Big]\,.
\end{equation}
This action leads to a coupled system of equations of motion,
\begin{equation}
\label{e:nonlinear}
\begin{aligned}
\partial_r \left[ \sqrt{-\hat{g}}e^{-\phi} \hat{g}^{00}\hat{g}^{rr} \partial_r A_0^{(3)} \right] &= 
4 (A_3^{(1)})^2 A_0^{(3)}\,\sqrt{-\hat{g}}e^{-\phi}\hat{g}^{00}\hat{g}^{33}\,,\\[1ex]
\partial_r \left[ \sqrt{-\hat{g}}e^{-\phi} \hat{g}^{33}\hat{g}^{rr} \partial_r A_3^{(1)} \right] &= 
4 (A_0^{(3)})^2 A_3^{(1)}\,\sqrt{-\hat{g}}e^{-\phi}\hat{g}^{00}\hat{g}^{33}\,,
\end{aligned}
\end{equation}
These are to be supplemented with boundary conditions at infinity,
which take the form
\begin{equation}
A_0^{(3)} = \mu + {\cal O}(r)\,,\quad
A_3^{(1)} = {\cal O}(r)\,.
\end{equation}
and boundary conditions at the horizon, which follow from the
requirement that \mbox{$A^{(3)}_0(r_T) = 0$}.  One typically finds
that for values of the chemical potential below a critical one,
$A_3^{(1)}$ is identically zero and the condensate is purely
abelian. For chemical potentials larger than \mbox{$\mu=\mu_c$},
the~$A_3^{(1)}$ component turns on, making the condensate
non-abelian. These solutions are parameterised by particular
combinations of values of~$\partial_r A_0^{(3)}$ and $A_3^{(1)}$ at
the horizon. If one keeps increasing~$\mu$ substantially beyond the
critical value, new branches of solutions occur. These typically have
a larger energy and hence are expected to be unstable. An example of
the structure of these solutions is given in
figure~\ref{f:D3D7_branches}.

\subsection{Overview of the models}

We will consider both three-dimensional and four-dimensional
superconductors of the type described above. For the three-dimensional
models will will focus on the $\text{AdS}_4$ black hole and the
non-conformal D2/D6 intersection.\footnote{The non-abelian condensate
  in the $\text{AdS}_4$ black hole background was previously
  constructed in~\cite{Gubser:2008zu}. The D2/D6 system allows, to
  some limited extent, an analysis which includes back-reaction
  effects~\cite{Erdmenger:2004dk}, and it would be interesting to
  investigate how these influence the results obtained in the present
  section.} Our main emphasis in the context of four-dimensional
superconducting systems will be the one obtained from the
Sakai-Sugimoto model~\cite{Sakai:2004cn}. However, in order to
appreciate the variety of ways in which this model is a non-trivial
superconductor, we will here also review some known facts about the
D3/D7 superconductor, and extend them with several new results.

A Frobenius analysis of the equations near infinity reveals that the
asymptotic behaviour of $A_0^{(3)}$ and $A_3^{(1)}$ is
\begin{equation}
\begin{aligned}
\text{AdS}_4: &\quad & A_0^{(3)} &= \mu - \rho r + \cdots \, ,\quad &
                     A_3^{(1)} &= \mu' + \rho' r + \cdots \, ,\\[1ex]
\text{D2/D6}: &\quad & A_0^{(3)} &= \mu - \rho r^2 + \cdots \, ,\quad &
                     A_3^{(1)} &= \mu' + \rho' r^2 + \cdots \, ,\\[1ex]
\text{D3/D7}: &\quad& A_0^{(3)} &= \mu - \rho r^2 + \cdots \, ,\quad &
                     A_3^{(1)} &= \mu' + \rho' r^2 + \cdots \, ,\\[1ex]
\text{D4/D8}: &\quad& A_0^{(3)} &= \mu - \rho\, r^{3/2}+ \cdots\,,\quad &
                     A_3^{(1)} &= \mu' + \rho'\, r^{3/2} + \cdots\,.
\end{aligned}
\end{equation}

In all of the cases we will be interested in solutions for which the
chemical potential for $A_0^{(3)}$ is turned on, while it is set to
zero for $A_3^{(1)}$; in other words we are looking for solutions
which satisfy the boundary conditions $\mu\neq 0$ and $\mu'=0$.  As
usual we also require that, for regularity of the solution, at the
horizon $A_0^{(3)}(r_T)=0$. The equations of motion then imply that
the near-horizon behaviour of the fields is (for all cases)
\begin{equation}
 A_0^{(3)} = A' (r -r_T) +\cdots\,,\qquad A_3^{(1)} = B + \cdots \,.
\end{equation}
By performing a two-dimensional scan through the parameter space
spanned by $A'$ and $B$ and selecting those values for which the
solutions to~\eqref{e:nonlinear} satisfy the boundary condition
that~$\mu'=0$, we arrive at the condensate solutions.

It is convenient to express these solutions in terms of a
dimensionless condensate~$\hat{\rho}$, given for the various systems
by
\begin{equation}
\begin{aligned}
\text{AdS}_4: & \quad \hat{\rho} = \rho/\mu^2\,,\\[1ex]
\text{D2/D6}: & \quad \hat{\rho} = \rho/\mu^{7/3}\,,\\[1ex]
\text{D3/D7}: & \quad \hat{\rho} = \rho/\mu^3\,,\\[1ex]
\text{D4/D8}: & \quad \hat{\rho} = \rho/\mu^4\,,
\end{aligned}
\end{equation}
(similar expressions hold for~$\rho'$). The advantage of
using~$\hat{\rho}$ and~$\hat{\rho}'$ is that these only depend on one
dimensionless quantity~$\mu/T$, not on the product~$TL$ (see 
appendix~\ref{s:scaling} for more details).

\begin{figure}[t]
\begin{center}
\includegraphics[width=.4\textwidth]{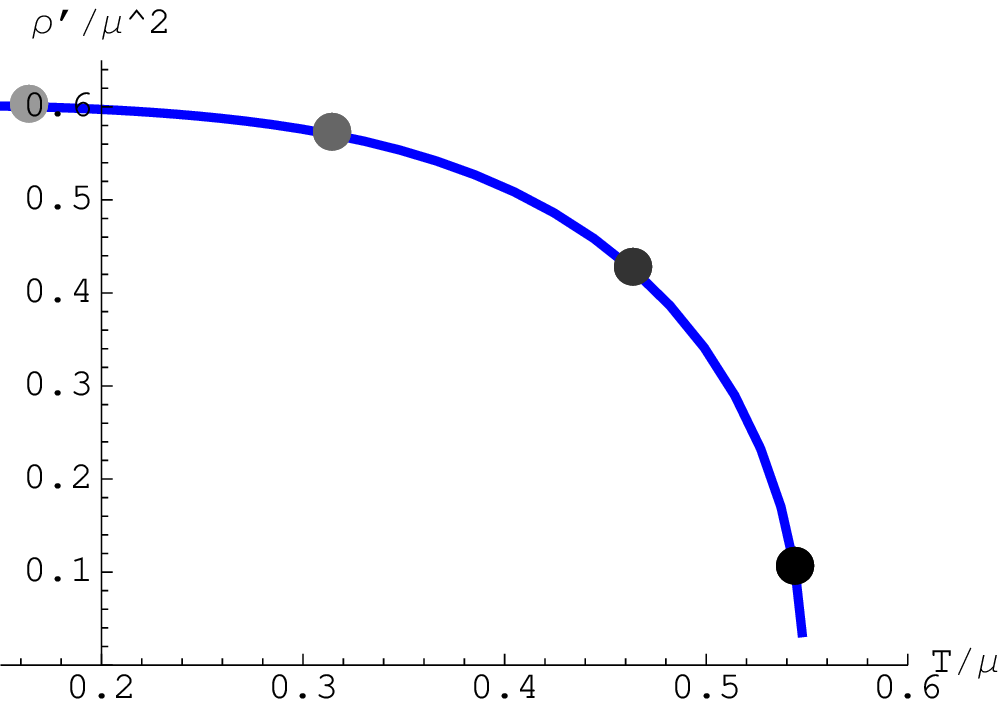}\qquad
\includegraphics[width=.4\textwidth]{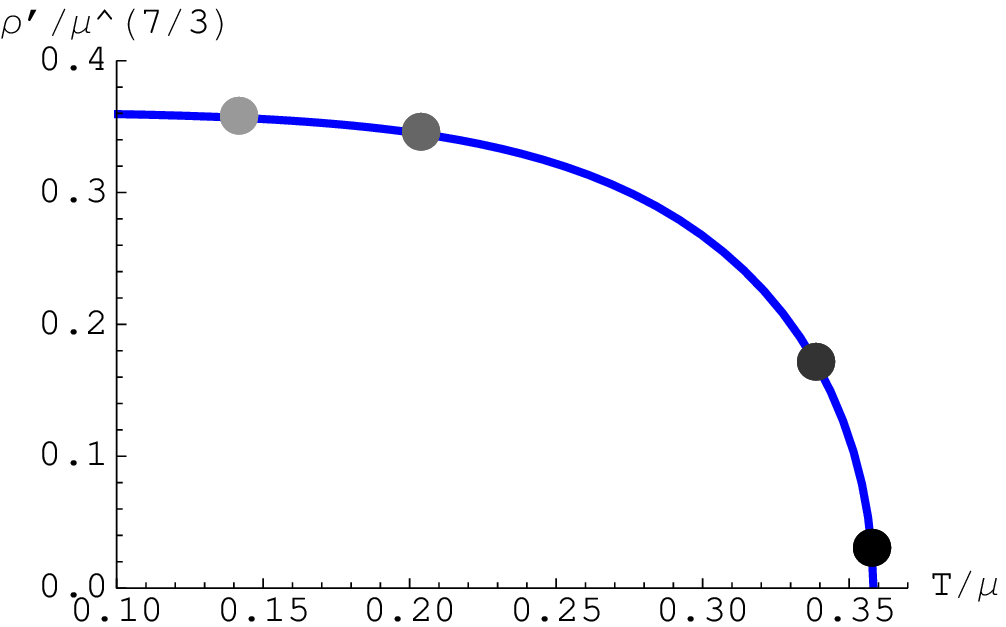}\\[2ex]
\includegraphics[width=.4\textwidth]{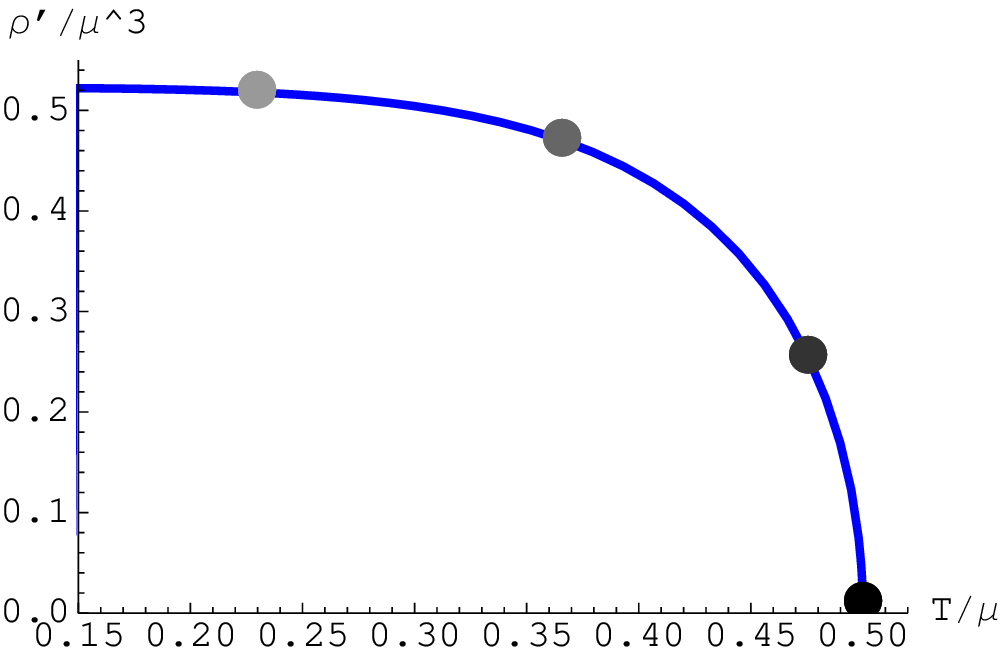}\qquad
\includegraphics[width=.4\textwidth]{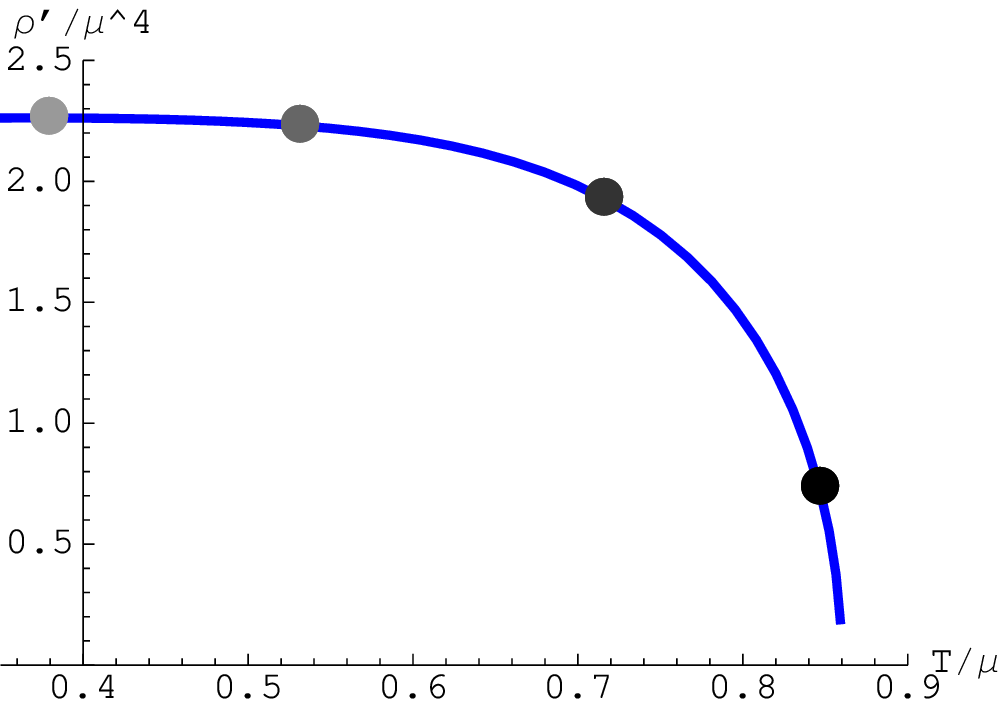}
\end{center}
\caption{The dimensionless condensates~$\hat{\rho}'$ for (left to
  right, top to bottom) the $\text{AdS}_4$ black hole, D2/D6, D3/D7
  and D4/D8 systems, as a function of the dimensionless ratio~$T/\mu$
  (or to be numerically precise, $r_T^{(p-5)/2}/\mu$). The dots indicate
  selected states for which further properties will be analysed in
  section~\protect\ref{s:EM}. \label{f:cases}}
\end{figure}

Plots of the dimensionless condensate~$\hat{\rho}'$ versus the
ratio~$T/\mu$ are given in figure~\ref{f:cases}.  We see that at fixed
temperature the condensate is nonvanishing when $\mu\leq \mu_c$, while
$\hat{\rho}'=0$ for $\mu>\mu_c$. In other words, the system undergoes a
second-order phase transition as the chemical potential is changed. By
numerically fitting the curves we see that the critical coefficient
$\alpha$, which determines the scaling behaviour of the condensate in the
vicinity of the critical point according to $\hat{\rho} \sim
(T/\mu_c-T/\mu)^\alpha$, is $\alpha=1/2$, in agreement with
Landau-Ginzburg theory.

\begin{figure}[t]
\begin{center}
\includegraphics[width=.6\textwidth]{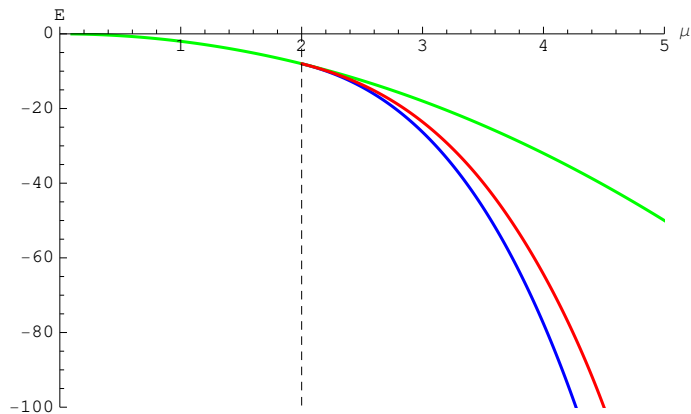}
\end{center}
\caption{Comparison of the energy of the two non-abelian condensates
  and the abelian one, for the D3/D7 system, as a function of the
  chemical potential. The ``three-component
  condensate''~\protect\eqref{e:altcond} (middle curve) always has a
  larger energy than the ``two-component
  condensate''~\protect\eqref{e:lowestEcondensate} (lower curve), and both have,
  for~$\mu > \mu_c \approx 2.0$, a lower energy than the abelian one
  (upper curve). The other systems analysed in this paper exhibit
  similar behaviour.\label{f:D3D7condensate_comparison}}
\end{figure}

For the alternative condensate given in~\eqref{e:altcond},
which is governed by the equations
\begin{equation}
\label{e:nonlinear2}
\begin{aligned}
\partial_r \left[ \sqrt{-\hat{g}}e^{-\phi} \hat{g}^{00}\hat{g}^{rr} \partial_r A_0^{(3)} \right] &=
8 A^2 A_0^{(3)}\,\sqrt{-\hat{g}} \hat{g}^{00}  \hat{g}^{33} e^{-\phi} \,,\\[1ex]
\partial_r \left[\sqrt{-\hat{g}}e^{-\phi} \hat{g}^{33}\hat{g}^{rr} \partial_r A \right] &= 
4 \Big[ (A_0^{(3)})^2\hat{g}^{00} - A^2\hat{g}^{33}\Big] A\,\sqrt{-\hat{g}} \hat{g}^{33}e^{-\phi}\,,
\end{aligned}
\end{equation}
the density as a function of the inverse chemical potential behaves
similarly. However, the energy of this configuration, obtained by
evaluating~$E = - L$ with~$L$ given by~\eqref{e:generic_action} is
consistently higher. A comparison is displayed in
figure~\ref{f:D3D7condensate_comparison}. We will henceforth restrict
ourselves to the lowest-energy condensate, which is the true ground
state of the model.

\subsection{Comments on the zero temperature limit}
\label{s:zeroTSS}

The strict zero temperature limit of holographic superconductors
remains somewhat of a mystery. In principle, there are three different
scenarios. In the first one, one takes a naive zero temperature limit
of the black hole metric, and obtains a metric with a Poincar\'e
horizon. This is what happens in the D3/D7 system upon
taking~$r_T\rightarrow 0$.  The resulting system suffers from the fact
that a condensate now becomes singular, even when using the full DBI
action.

Another possibility is that the limit should be taken in such a way
that an extremal horizon is obtained. Concrete examples of this type
have not yet been found, but they would presumably avoid the
triviality of the Poincar\'e horizon case discussed above. An attempt
to analyse the zero-temperature situation was made
in~\cite{Gubser:2008wz} by constructing a solution interpolating
between two different $\text{AdS}_4$ metrics. However, because of the
conformal symmetry in the infrared, the solution found there exhibits
a power-law scaling of the conductivity for small frequency, and hence
no gap~\cite{Gubser:2008pf}.

There remains on further option, which seems more natural in e.g.~the
Sakai-Sugimoto model. In contrast to e.g.~the $\text{AdS}_4$ black
holes used in many studies of holographic superconductors, the
Sakai-Sugimoto model has a well-understood limit to zero
temperature. For sufficiently low temperature, the system undergoes a
first order phase transition~\cite{Aharony:2006da}. One then obtains a
D4/D8 system in which there is no horizon on the brane, and the
D8-branes which were previously intersecting the horizon are now
connected in a smooth way. This system admits non-abelian condensates
(as analysed in~\cite{Aharony:2007uu}). The absence of a horizon
automatically implies that there is no dissipation in the system.

\section{Thermodynamic properties}
\label{s:thermo}
\subsection{Free energy and specific heat}
\label{s:free_energy}

Let us now turn to an analysis of some of the thermodynamic properties
of the superconductors. We will first present the results for the
holographic models and then we will compare these with real world
superconductors in section~\ref{s:realworld_cv}.

There has been some confusion in the literature as to the parameters
which should be kept fixed when computing the specific heat. From the
point of view of an experimental setup, keeping the density~$\rho$
constant is the most natural. This type of computation has been done
for the abelian condensate of the D3/D7 case, using a full DBI
analysis, in~\cite{Karch:2008fa}. Alternatively, one could consider
keeping the chemical potential~$\mu$ constant, which is a natural
thing to do from the point of view of some holographic setups. A
computation of this type can be found in
e.g.~\cite{Kulaxizi:2008jx}. Depending on what is kept fixed, the
results can be quite different in as far as the dependence on the
temperature is concerned. We will here elucidate these differences and
then continue to show how the results change when one considers a
non-abelian condensate instead.

The main technical ingredient in the computation of the temperature
dependence of the free energy is the scaling symmetry discussed in
section~\ref{s:scaling}. It allows us to study the temperature
dependence by analysing the dependence on the chemical potential~$\mu$
at fixed temperature. By writing the Euclidean action in dimensionless
coordinates and dimensionless fields~$\tilde{A}_0$ and $\tilde{A}_3$,
we find that
\begin{equation}
\label{e:SSt}
\begin{aligned}
\text{AdS}_4:& \qquad & S_E(T) &= (TL)^2\,\times \tilde{S}_E(\mu/T)\,,\\[1ex]
\text{D2/D6}:& \qquad & S_E(T) &= (TL)^{\frac{7}{3}}\,\times \tilde{S}_E(\mu/T)\,,\\[1ex]
\text{D3/D7}:& \qquad & S_E(T) &= (TL)^3\,\times \tilde{S}_E(\mu/T)\,,\\[1ex]
\text{D4/D8}:& \qquad & S_E(T) &= (TL)^4\,\times \tilde{S}_E(\mu/T)\,,
\end{aligned}
\end{equation}
In writing these factors we have taken into account that the integral
over the Euclidean time circle produces a factor of~$1/T$.

For the computation of the heat capacity we now use the free energy~$F
= T S_E$ together with
\begin{equation}
\label{e:cvdef}
c_v = T \frac{\partial s}{\partial T}\qquad \text{with}\qquad
s = -\frac{\partial F}{\partial T}\,,
\end{equation}
where~$s$ is the entropy density, and we have suppressed volume
factors.  Using the form~\eqref{e:SSt} it is easy to compute
quantities like the free energy or the specific heat at fixed chemical
potential~$\mu$. In order to compute at fixed density~$\rho$, one has to
express~$\mu/T$ in terms of~$\hat{\rho}$ using condensate curves
like those of figure~\ref{f:cases}.

\begin{figure}[t]
\includegraphics[width=.45\textwidth]{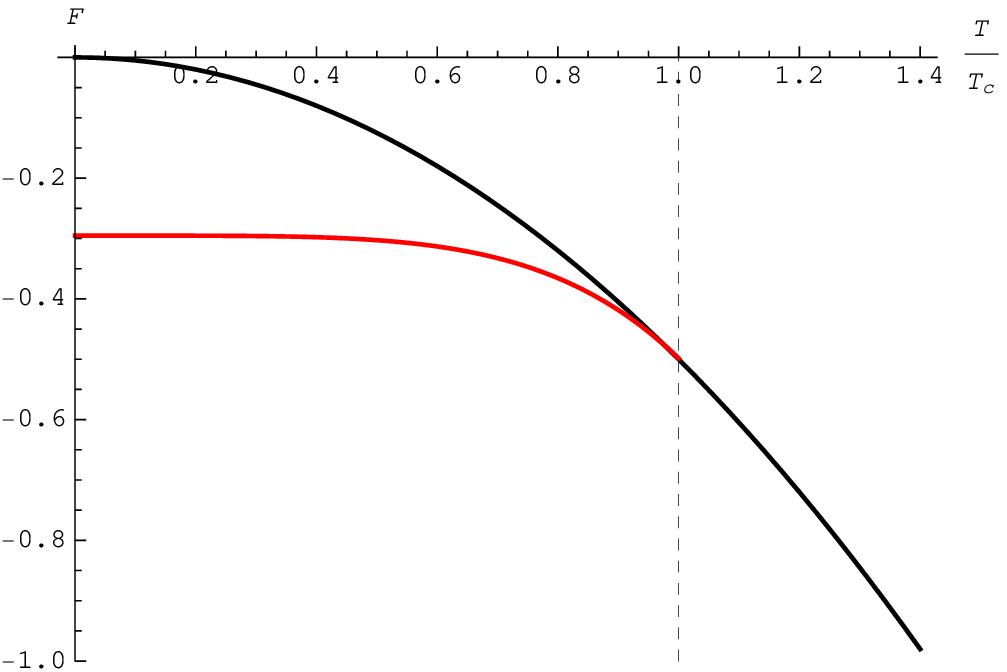}\qquad
\includegraphics[width=.45\textwidth]{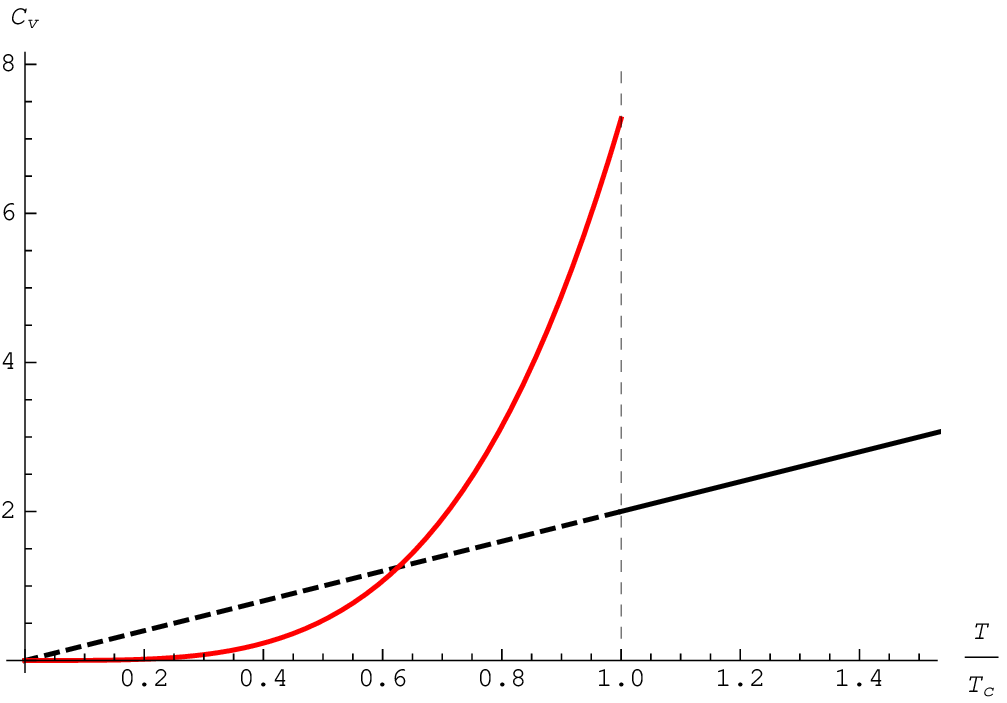}
\caption{Free energies and specific heat for the D3/D7 system, as a function of temperature, for fixed
  chemical potential~$\mu$ (and hence with~$\rho=\rho(T)$). The black
  curve denotes the abelian condensate with only~$A_0$ non-zero,
  while the red curve denotes the non-abelian condensate.\label{f:Fs}}
\end{figure}

Figure~\ref{f:Fs} shows curves for the free energy~$F(T)$ at constant
$\mu$; the curves at constant~$\rho$ are qualitatively similar. After
normalising \mbox{$F(T=0)=0$}, a double-logarithmic plot (not shown
here) leads to practically straight curves, which suggests a fit of
the free energy to a function of the form~$a + b T^c$ (as opposed to
something involving exponentials of the temperature). We will comment
below on how this connects to real-world superconductors. From the
fit of the free energy we then obtain~$c_v$ by applying~\eqref{e:cvdef} to
the fit, and arrive at the following numerical estimates in the normal
and superconducting phases\footnote{Numerically it is more reliable to
  fit~$F$ rather than~$c_v$. The ansatz is consistent with the fact
  that~$c_v$ should vanish as~$T\rightarrow 0$, as required by the
  third law of thermodynamics.},
\begin{equation}
\label{listofCmu}
\begin{tabular}{lcccc}
                & normal phase &
                  \multicolumn{2}{c}{superconducting} \\
                & $c_v(\mu=\text{ct.})$ &
                  $c_v(\mu=\text{ct.})$ & $c_v(\rho=\text{ct.})$ \\[1ex]
$\text{AdS}_4$:\quad & $T^{0}$ & $T^{2.7}$ & $T^{2.9}$ \\[1ex]
D2/D6:         \quad & $T^{\frac{1}{3}}$   & $T^{3.0}$ & $T^{3.2}$ \\[1ex]
D3/D7:         \quad & $T$     & $T^{3.8}$ & $T^{3.9}$ \\[1ex] 
D4/D8:         \quad & $T^{2}$ & $T^{5.5}$    & $T^{5.2}$ 
\end{tabular}
\end{equation}
As explained below, the results for the free energy in the abelian
phase at constant~$\rho$ are unreliable in the Yang-Mills truncation,
and we have therefore not indicated values for~$c_v(\rho=\text{const.})$
in the normal phase.  The systematic error bars on the numerical
exponents receive contributions from various stages of the
construction and should thus only be used as indicators about the
numbers of degrees of freedom.

\begin{figure}[t]
\includegraphics[width=.32\textwidth]{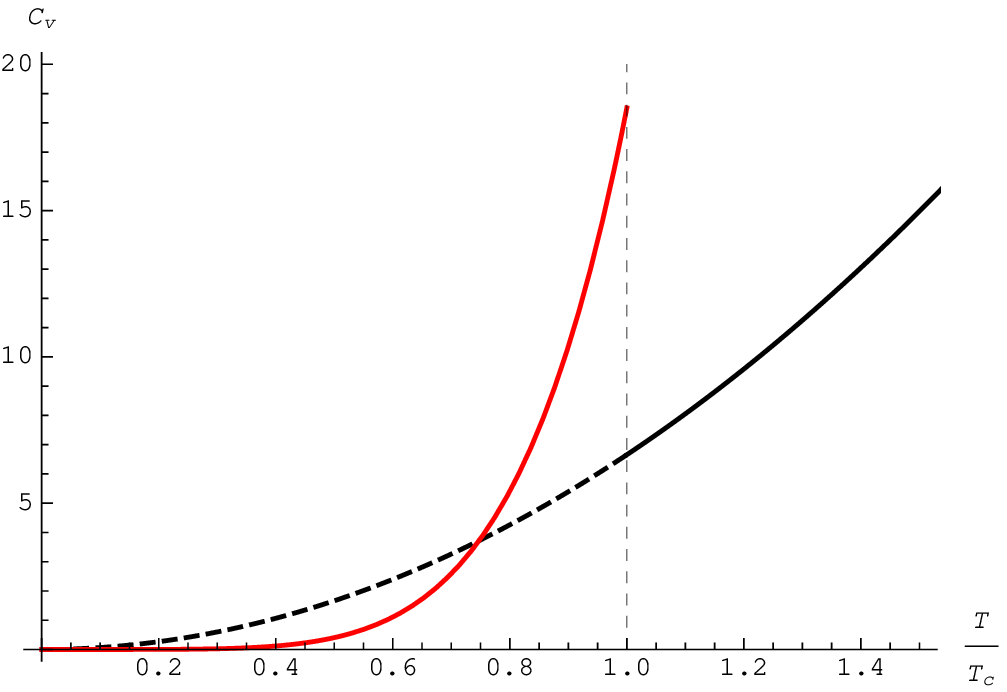}\;\;
\includegraphics[width=.32\textwidth]{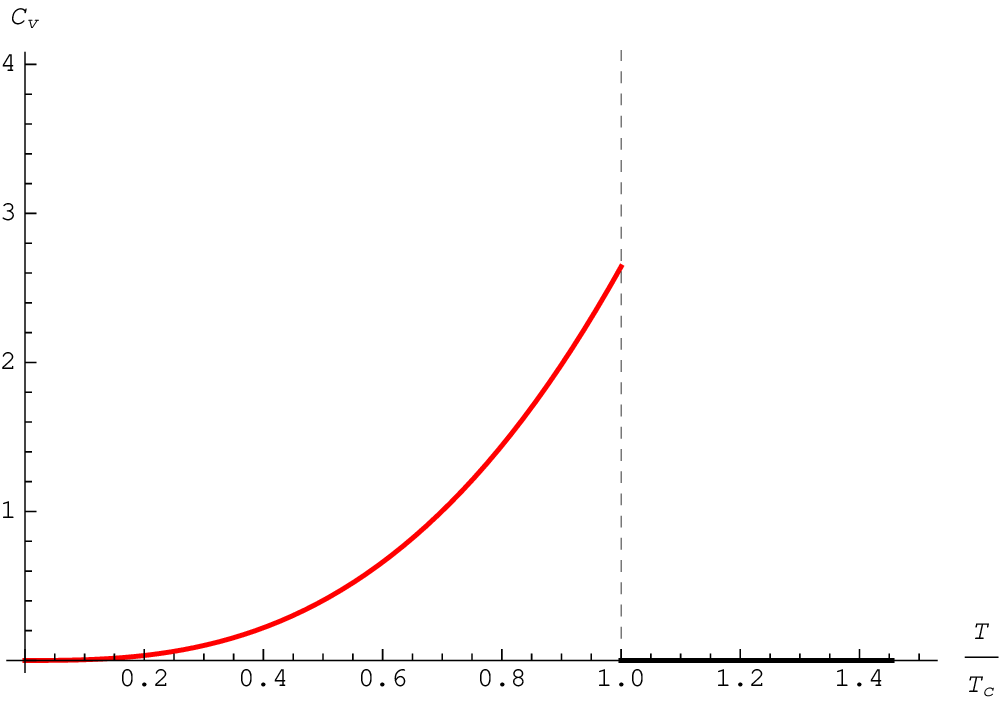}\;\;
\includegraphics[width=.32\textwidth]{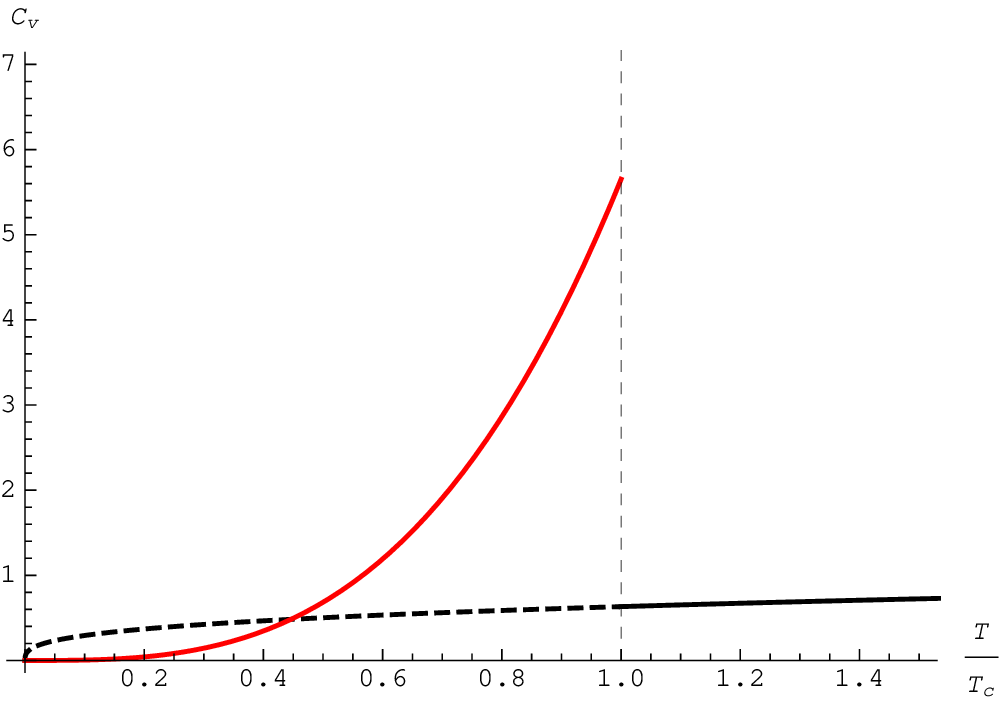}
\caption{Specific heat for the D4/D8, $\text{AdS}_4$ and D2/D6 systems
  respectively; curves as in figure~\protect\ref{f:Fs}. The free
  energy plots of these systems are similar to figure~\protect\ref{f:Fs}a, the
  only difference being the exponents of the curves.\label{f:FsD4D8}}
\end{figure}

The only previous result about the specific heat of a non-abelian
condensate is, as far as we are aware, given in the work
of~\cite{Ammon:2009fe}. These authors claim that
\mbox{$c_v(\rho=\text{const}.)\sim T^3$} for the D3/D7 system
(analysed using the non-abelian DBI action in two different
prescriptions). Our computations for D3/D7 lead to a somewhat larger
exponent: they also contain a factor~$T^3$, arising from the overall
prefactor in~\eqref{e:SSt}, but we find an additional temperature
dependence through~$\tilde{S}(\mu/T)$. This difference may be
a consequence of the DBI corrections to the Yang-Mills action, but we
did not check the results of~\cite{Ammon:2009fe}.

The most important conclusion to draw from our analysis -- more
important than the actual numbers -- is the fact that we find no
evidence for exponential suppression of the specific heat in the
superconducting phase. As we will see, this is in sharp contrast with 
BCS-like superconductors. The second important result is that the
non-abelian systems exhibit a jump in the specific heat at the phase
transition. We will discuss the relevance of these findings to real
world superconductors in \S\ref{s:realworld_cv}.

\subsection{Limitations of the Yang-Mills approximation}
\label{s:limitations}

The result for the specific heat at constant \emph{density} for the
abelian condensate (i.e.~above $T_c$), which were obtained from our
Yang-Mills analysis, differ substantially from the ones obtained using
a DBI analysis~\cite{Karch:2008fa}. However, we have also seen that
computing the specific heat above $T_c$ at constant \emph{chemical
  potential}, gives a much better agreement with the DBI
computation~\cite{Kulaxizi:2008jx}. The reason for this seems directly
related to the behaviour of the corresponding two gauge field
solutions in the zero-temperature limit, as the one at constant
chemical potential admits a smooth $T\rightarrow 0$ limit while the
solution at fixed density does not.

To see this explicitly, consider the abelian solution to the
Yang-Mills equations of motion, for instance for the D4/D8 model. It
reads
\begin{equation}
\partial_r A_0 = \rho\, r^{1/2}\,.
\end{equation}
The chemical potential and the free energy are (making use
of~\eqref{e:HawkingT}, i.e.~$T = 3/(4\pi\sqrt{r_T})$) 
\begin{equation}
\begin{aligned}\mu &= \int_0^{r_T} \partial_r A_0 
     = \int_{0}^{r_T} {\rm d}r\, \rho r^{1/2} 
     = \frac{2}{3} \rho \left(\frac{4}{3}\pi T\right)^{-3}\,,\\[1ex]
F   &= \int_0^{r_T}\! \sqrt{-g} e^{-\phi} g^{rr} g^{00} (\partial_r A_0)^2 = 
\int_{0}^{r_T}\!{\rm d}r\, \rho^2 r^{1/2} = \rho \mu\,.
\end{aligned}
\end{equation}
The latter expression yields~$\partial F/\partial\mu = \rho$,
confirming that~$\rho$ is the density. Clearly, if we keep~$\rho$
fixed and take the limit~$T\rightarrow 0$, the gauge field becomes
singular at the horizon~$r=r_T$. However, if we eliminate~$\rho$ in
favour of~$\mu$, then the gauge field becomes \mbox{$\partial_r A_0
  =\frac{3}{2}\mu \left(\frac{4}{3}\pi T\right)^3
  r^{1/2}$}. At~$r=r_T$ this is regular for~$T\rightarrow 0$. A
similar analysis holds for the other models analysed in this paper. In
contrast, the non-abelian solutions  always seem to be regular in
the~$T\rightarrow 0$ limit.

\subsection{Comments on connections with real-world superconductors}
\label{s:realworld_cv}

Knowledge of the specific heat in solid state physics is of central
importance, as it shows what are the relevant degrees of freedom, and
how these vary with temperature. In addition, this quantity is in a
sense ``simpler'' and more robust than e.g.~the resistivity, since, as
a scalar quantity, it is insensitive to the anisotropic properties of
the material.  For standard metals the specific heat~$c_v$ behaves as
\begin{equation}
\label{Cstandard}
c_v \equiv c_v^{\text{el}} + c_v^{\text{ph}} = \gamma T + \beta T^3\,.
\end{equation}
The first term, proportional to the Sommerfeld constant~$\gamma$,
originates from an electronic contribution and the second term
originates from lattice phonons. While the lattice contribution to the
specific heat is unchanged for superconductors, the electronic part
varies significantly, depending on the type of superconductor~(see
e.g.~\cite{Tari:2003a}).

\begin{figure}[t]
\begin{center}
\includegraphics[width=.42\textwidth]{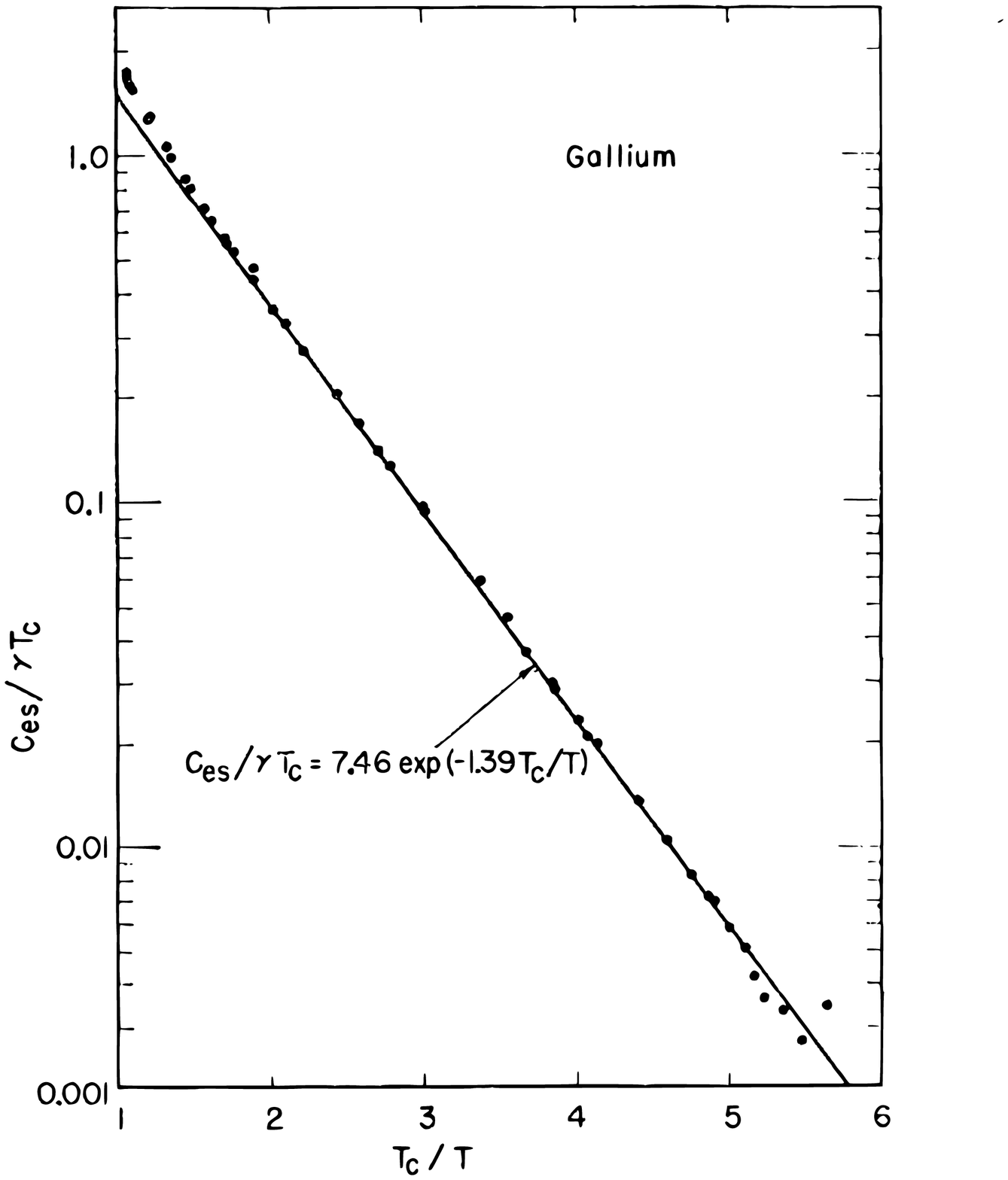}\quad
\raisebox{1.2ex}{\includegraphics[width=.45\textwidth]{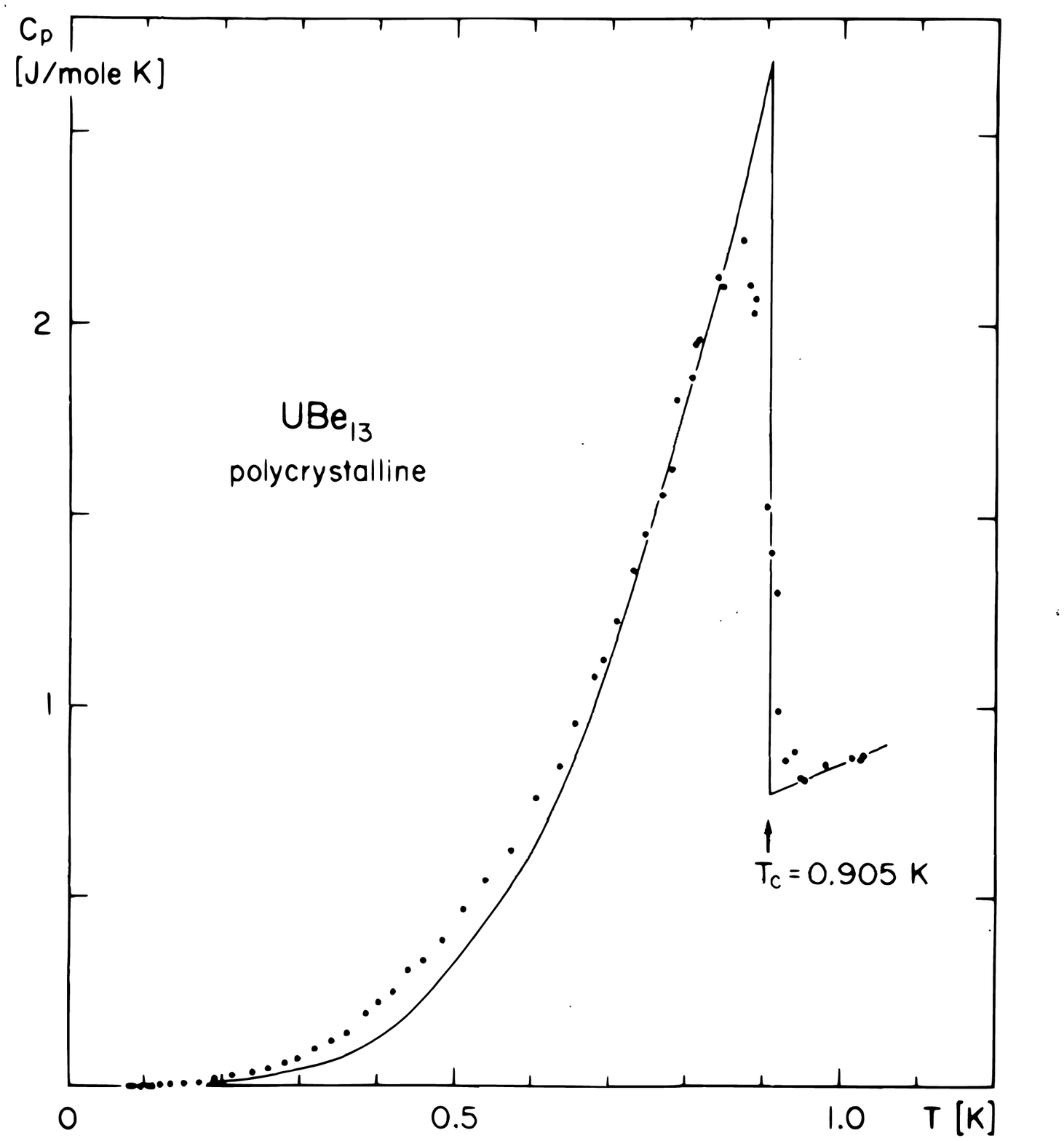}}
\end{center}
\caption{Specific heat of the low-$T_c$ BCS superconductor gallium (left)
  and the heavy-fermion compound $\text{UBe}_{13}$ (right) in their
  superconducting phases. Figures
  taken from~\cite{Phillips:1964a} and~\cite{Ott:1984a}
  respectively. \label{f:gallium_UBe13_C}}
\end{figure}

The weakly coupled superconductors are successfully described with BCS
theory, which predicts that the specific heat is exponentially
suppressed for temperatures $T<T_c$, i.e.
\begin{equation}
\label{BCSC}
\frac{c_v}{\gamma T_c} = A e^{- B \frac{T_c}{T}}\,.
\end{equation}
Here $\gamma$ is the Sommerfeld constant in the normal phase, and $A$
and $B$ are constants that depend on the temperature
interval. Therefore, for sufficiently low temperatures the graph
$\ln(C/\gamma T_c)$ against $T_c/T$ is a straight line, which is also
experimentally observed (see the left panel of
figure~\ref{f:gallium_UBe13_C}). The exponential suppression of the
specific heat is a natural consequence of the presence of a gap, since
when $T<T_c/2$, the energy necessary to break a Cooper pair is $\sim 2
\Delta(0)$, and the number of broken pairs is $\sim e^{- 2
  \Delta(0)/k_B T}$.  Another important feature of low $T_c$
superconductors is that the specific heat is \emph{discontinuous} at
$T=T_c$ since at this temperature it is no longer necessary to put
energy into breaking of Cooper pairs. BCS theory predicts a universal
number for this jump, $\Delta c(T_c)/\gamma T_c = 1.43$, which is
indeed observed in a large number of low-$T_c$ superconductors.

Another class of superconductors are the \emph{strongly coupled}
superconductors which include for example high-$T_c$ superconductors
and heavy fermion compounds. Most of these materials are
experimentally found to involve higher-spin electric carriers. From
this perspective one might expect that these materials should be more
similar to non-abelian (spin one) condensates considered in this paper.
Heavy fermion compounds are very different from BCS-like materials
when cooled down to the superconducting phase. They have a specific
heat below $T_c$ which is well-approximated with a power law
behaviour~(see e.g.~\cite{Riseborough:2008a}),
\begin{equation}
\label{HFsuperCv}
\text{heavy fermions at $T<T_c$}:\qquad c_v = \gamma_0 T + \kappa T^n\,,
\end{equation}
where $n \sim 2-3$, and $\gamma_0$ and $\kappa$ are constants. In
other words, due to the unusual properties of the gap, the specific
heat is no longer exponentially suppressed when $T<T_c$. The specific
heat jump at $T=T_c$, which is observed for the weakly coupled materials, does
exist for the strongly coupled superconductors as well, although the
normalised value of the jump is, for many systems, much larger than
the value~$1.43$ predicted by BCS theory (see the right panel of
figure~\ref{f:gallium_UBe13_C}).

When heated above~$T=T_c$, a large number of heavy fermion compounds
behave as a (heavy) fermion liquid, i.e.~with specific heat $c/T \sim
\gamma(T)= \text{const}.$ . However, some of them (for example
$\text{UBe}_{13}$) have an unusual behaviour and their specific heat
is well-described by
\begin{equation} 
\label{HFCaboveTc}
\text{heavy fermions at $T>T_c$}:\quad
c_v \sim \frac{T}{T_0} \ln\left(\frac{T_0}{T}\right) \quad
\text{or} \quad
c_v \sim T^{\lambda} \quad (\lambda \approx 0.7-0.8) \, .
\end{equation}

Keeping all this in mind we can try to compare some of our findings
with real superconductors. Firstly, we see that in the superconducting
phase, colourful holographic superconductors have power-law behaviour
of the specific heat, which is closer to the strongly coupled
superconductors, not the BCS superconductors. Secondly, all systems
exhibit a jump in the specific heat, in agreement with real-world
superconductors.

Finally, when $T>T_c$, in all cases the specific heat \emph{increases}
with temperature according to a power law. The precise value of the
positive exponents depends on details of the system, on whether we
work at fixed chemical potential or at fixed density (see the previous
section), and finally it depends on whether we work in the Yang-Mills
approximation (see table~\ref{listofCmu}) or with the full DBI
action~\cite{Karch:2008fa}.  From the experimental perspective the
specific heat at fixed $\rho$ is a more interesting quantity, and as
we have already explained the Yang-Mills approximation is in this case
not appropriate. To analyse the $T>T_c$ temperature range, we thus use
the results of~\cite{Karch:2008fa}. All cases considered in
\cite{Karch:2008fa} are conformal, such that the non-sphere part of
the D-brane worldvolume is $\text{AdS}_{p+2}$. They find that $c\sim
T^{2p}$. Hence none of the conformal cases seem to fit the metallic
behaviour of heavy fermion compounds. The specific heat at fixed $\mu$
seems more close to real-world superconductors, since in the
Yang-Mills approximation the D2/D6 system has exponent $1/3$,
resembling~\eqref{HFCaboveTc} and D3/D7 has metallic behaviour, while
in the DBI approximation the D4/D8 system exhibits metallic
behaviour~\cite{Kulaxizi:2008jx}.

In summary, from the perspective of the specific heat, most of the
colourful holographic superconductors show some qualitative
similarities with strongly coupled superconductors. In particular,
modulo subtleties discussed above, the D3/D7 system looks very similar
to e.g.~the $\text{UBe}_{13}$ heavy fermion superconductor.

\section{Electromagnetic properties}
\label{s:EM}

In this section we turn to a study of the response functions for the
superconducting condensates that were constructed in the previous
sections. In particular, we focus on the computations of AC and DC
conductivities. At the end we compare our holographic findings with
qualitative features of real-world superconductors.

\subsection{AC conductivity}

\afterpage{\clearpage
\begin{figure}[t]
\begin{center}
\includegraphics[width=.45\textwidth]{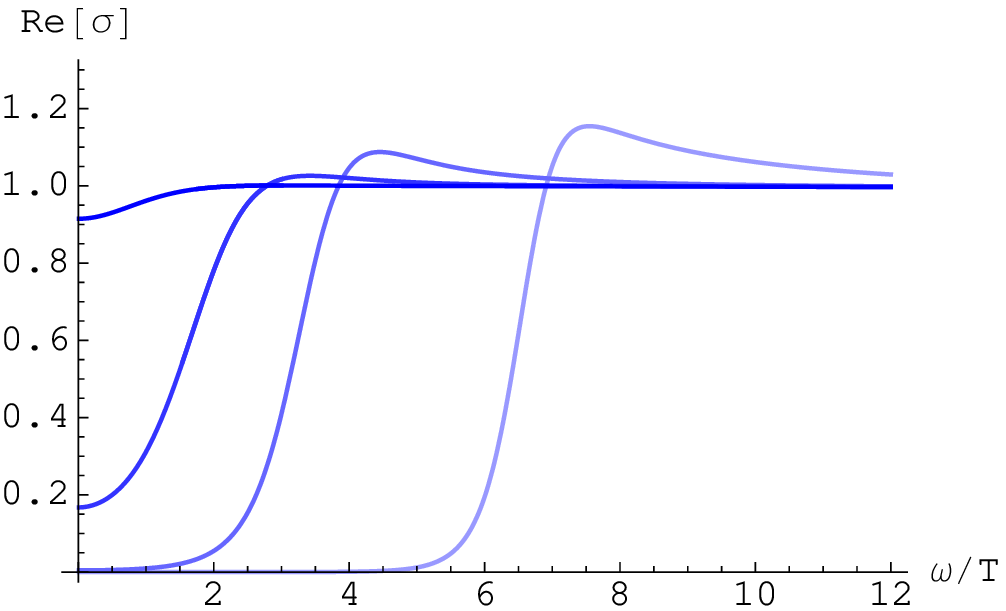}\quad
\includegraphics[width=.45\textwidth]{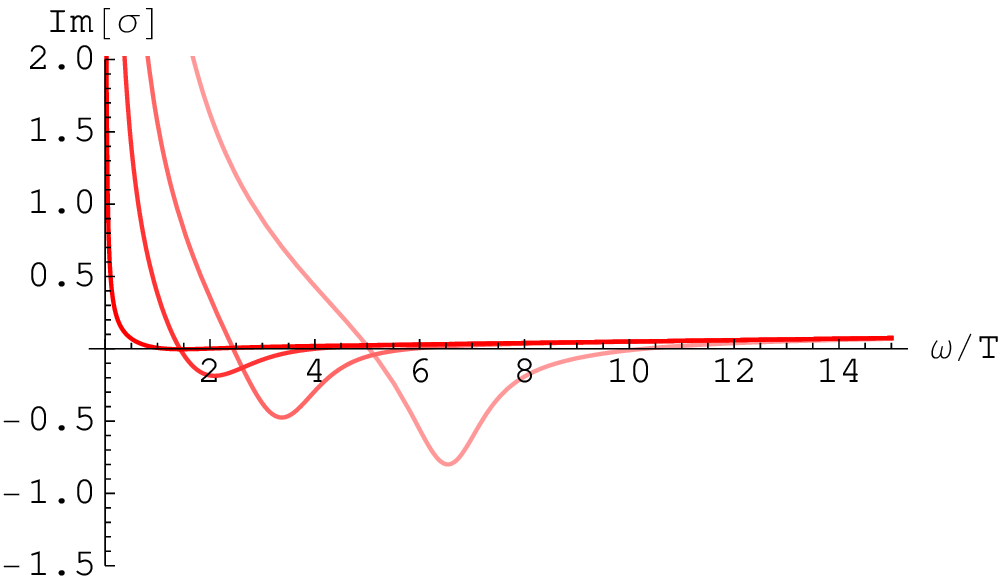}\\[4ex]
\includegraphics[width=.45\textwidth]{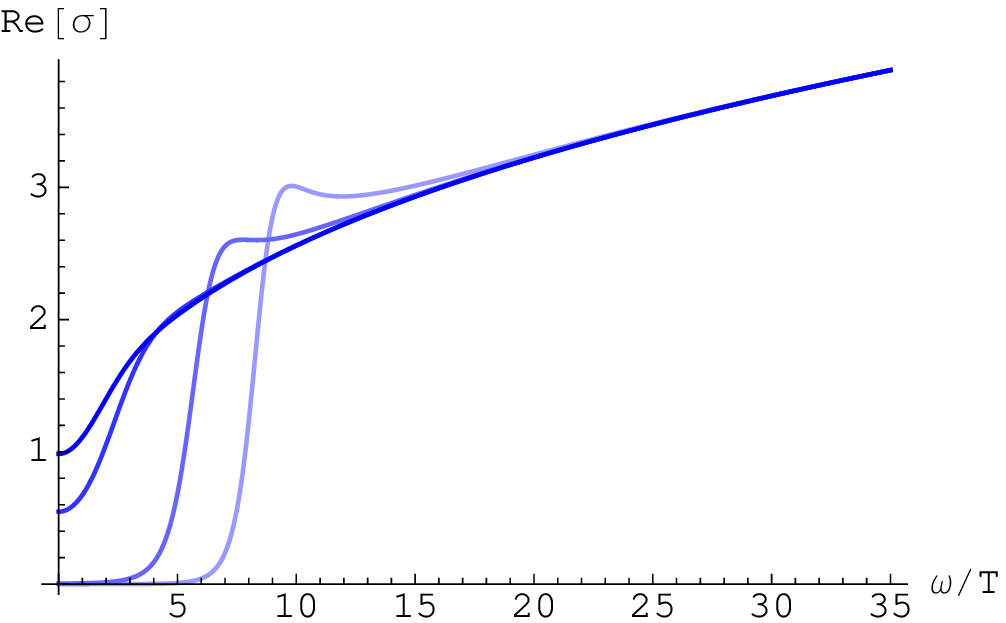}\quad
\includegraphics[width=.45\textwidth]{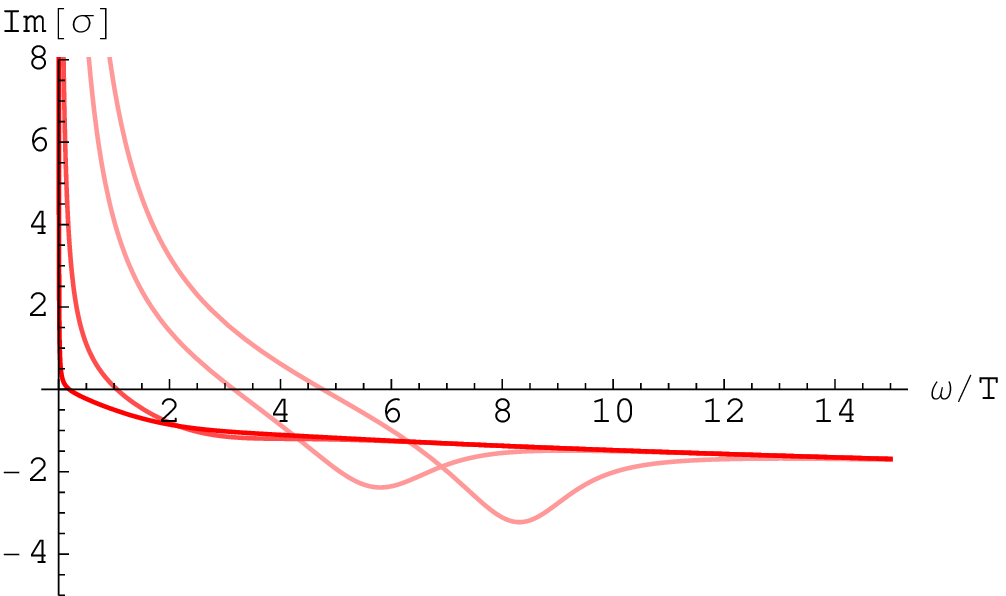}\\[4ex]
\includegraphics[width=.45\textwidth]{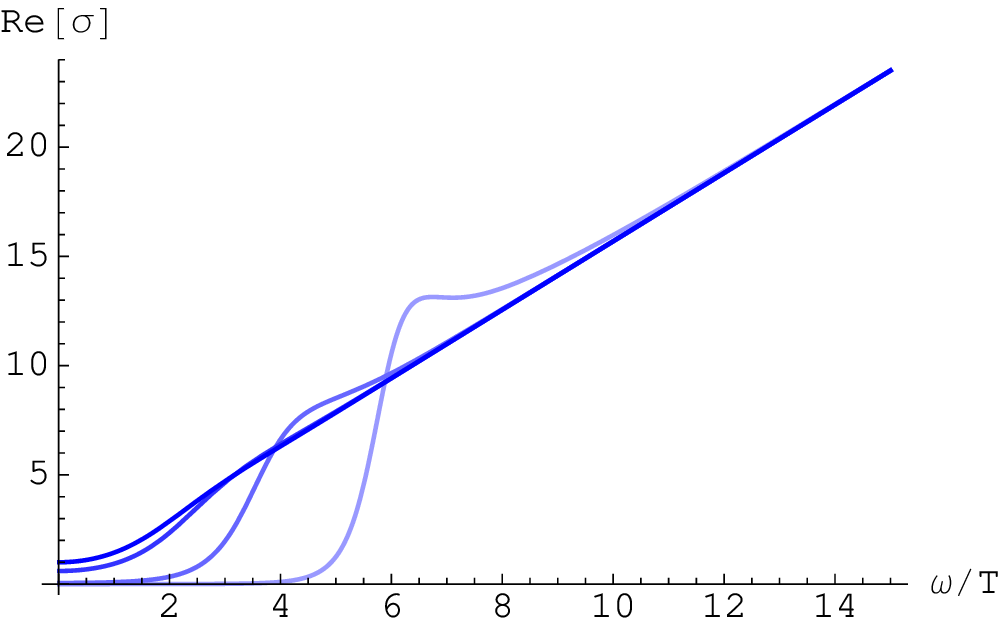}\quad
\includegraphics[width=.45\textwidth]{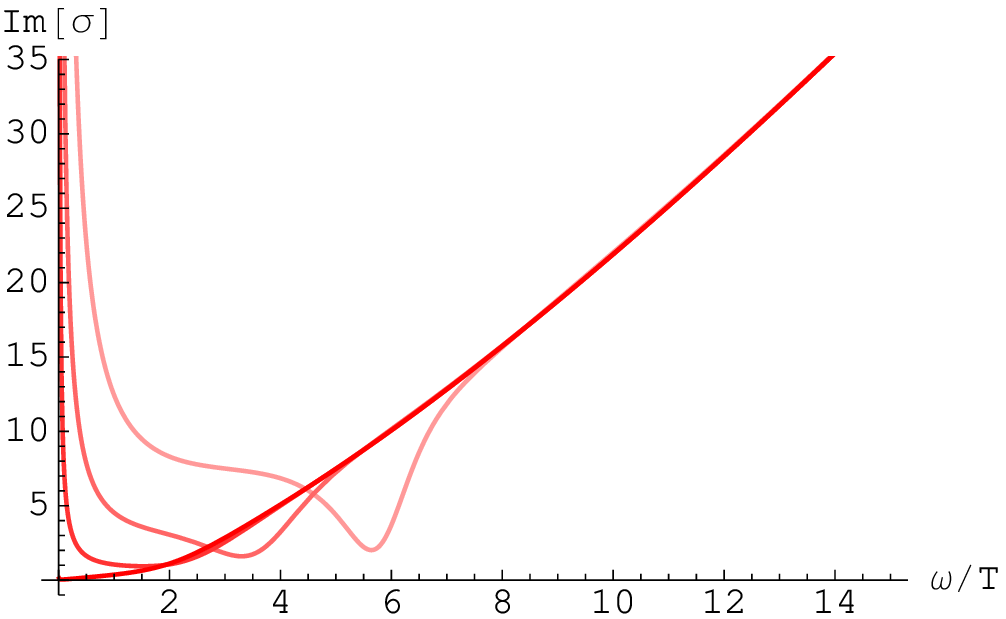}\\[4ex]
\includegraphics[width=.45\textwidth]{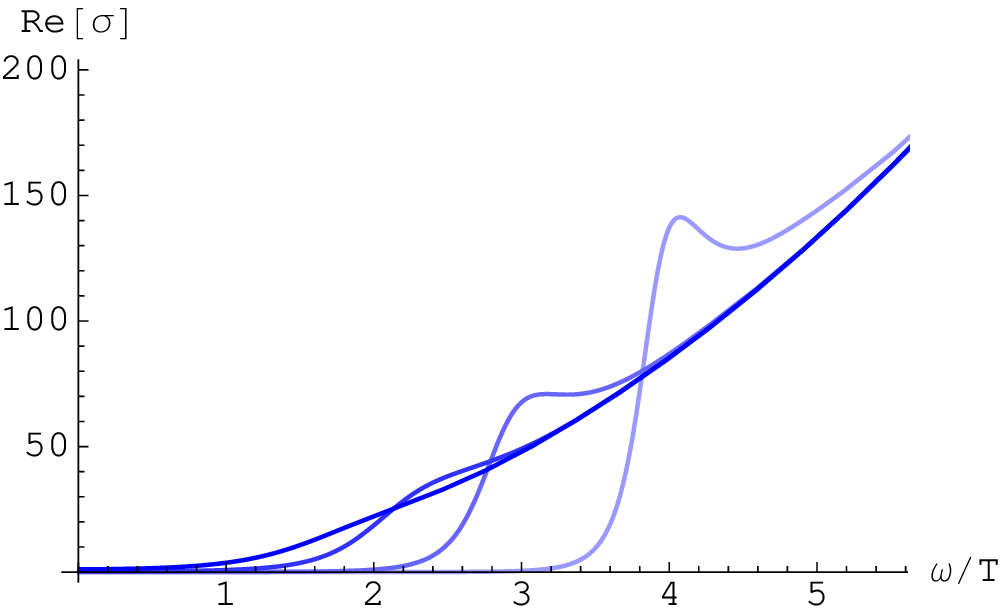}\quad
\includegraphics[width=.45\textwidth]{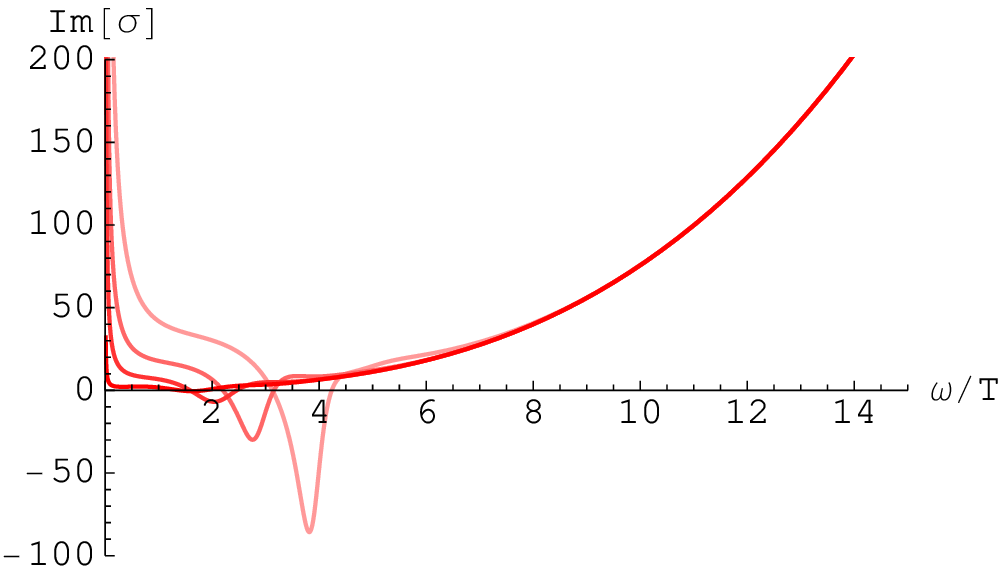}
\end{center}
\caption{Real and imaginary parts of the conductivity for the four
  systems (from top to bottom: $\text{AdS}_4$, D2/D6, D3/D7 and
  D4/D8), plotted for various values of the dimensionless chemical
  potential~$\mu/T$ (corresponding to the dots in
  figure~\protect\ref{f:cases}). \label{f:conductivities}}
\end{figure}\clearpage}

Once a condensate has been found, response functions in this
background can be obtained by analysing the Green function of the
gauge field fluctuations. For the
condensate~\eqref{e:lowestEcondensate}, one can in principle fluctuate
in the direction parallel to the condensate or orthogonal to
it~\cite{Gubser:2008wv}. Here we choose to consider only fluctuations
orthogonal to the condensate, i.e.~we look at the fluctuation modes in
the~$\psi \equiv A_2^{(3)}$ field. These fluctuation modes satisfy
\begin{equation}
\label{e:scla0}
\partial_r\Big[ \sqrt{-\hat{g}} e^{-\phi} \hat{g}^{rr}
  \hat{g}^{22} \partial_r\psi\Big] 
- 4 \sqrt{-\hat{g}} e^{-\phi} \hat{g}^{33} \hat{g}^{22} (A_3^{(1)})^2 \psi 
- \sqrt{-\hat{g}} e^{-\phi} \hat{g}^{00} \hat{g}^{22} \omega^2 \psi = 0\,.
\end{equation}
In particular, the Green function is obtained from the boundary data using
\begin{equation}
\label{e:Green}
G_R(\omega) = - \lim_{r\rightarrow 0}
     \sqrt{-\hat{g}} e^{-\phi} \hat{g}^{33} \hat{g}^{rr} \psi \partial_r\psi\,.
\end{equation}
and the AC conductivity then follows from
\begin{equation}
\label{e:conductivity_def}
\sigma(\omega) = \frac{G_R(\omega)}{i\omega}\,.
\end{equation}

The fluctuation equation~\eqref{e:scla0} exhibits properties just like
the condensate equations. Following an analysis similar to that for
condensate equations (see the appendix), we put it in the form
\begin{equation}
\label{ELfluct}
\partial_{\tilde{r}} \Big( h_3(\tilde{r}) \partial_{\tilde{r}}
\psi\Big) - 
\Big[
 4 (A_3^{(1)})^2 h_5(\tilde{r})
+ \omega^2 h_2(\tilde{r}) \Big] T^{-2} \psi = 0 \,.
\end{equation}
From this we conclude that the conductivity~$\sigma$ will be a
function of the two dimensionless parameters~$\tilde{\mu}$
and~$\hat{\rho}(\tilde{\mu})$ which appear in~\eqref{e:A0expdimless},
as well as the combination~$\omega/T$.

In the D3/D7 case, computing the conductivity requires some extra
care because of the presence of logarithmic terms in the asymptotic
expansion of the gauge field
fluctuations~\cite{Karch:2007pd,Horowitz:2008bn}. A perturbative
analysis yields
\begin{equation}
\psi = a_0\Big(1 - \tfrac{1}{2}\omega^2r^2 \log (\Lambda r) + \ldots \Big) + a_2 r^2 + \ldots\,.
\end{equation}
When evaluating the conductivity~\eqref{e:conductivity_def} this leads
to a logarithmically divergent term as well as a finite
contribution.\footnote{An analysis of the non-abelian condensate in
  the D3/D7 system has previously appeared in~\cite{Basu:2008bh} but
  that paper ignores the effect of the logarithmic terms in the
  fluctuation, as witnessed by e.g.~the different imaginary part of the
  conductivity.}  The logarithmically divergent term can be removed by
suitable addition of a holographic counterterm. 

We impose ingoing boundary condition on the fluctuation~$\psi$ at the
horizon, which means
\begin{equation}
\psi = (r_T-r)^{-i\alpha \omega} (1 + a_1 (r_T-r) + a_2 (r_T-r)^2 + \ldots)\quad\text{near~$r=r_T$}\,,
\end{equation}
where~$\alpha$ is a model-dependent real coefficient,
$\frac{1}{3}r_T$, $\frac{1}{5}r_T^{3/2}$, $\frac{1}{4}r_T$, and
$\frac{1}{3}\sqrt{r_T}$ for the $\text{AdS}_4$, D2/D6, D3/D7 and D4/D8
models respectively.  After shooting to the boundary at~$r=0$ we then
compute the retarded Green function from~\eqref{e:Green}. We have
analysed the real and imaginary parts of the conductivity for all four
systems; the results are displayed in
figure~\ref{f:conductivities}. 

There is a number of interesting features in both the three and
four-dimensional cases.  First, in comparison to holographic
superconductors obtained from $\text{AdS}_4$ black holes in
four-dimensional Abelian Higgs models, such as analysed
in~\cite{Horowitz:2008bn}, we see that for all other cases, in the
absence of a condensate, the real part of the conductivity does not
approach a constant. This is in particular true for the
two-dimensional, non-conformal D2/D6 system.  Second, we see that the
value of the chemical potential where a gap in the real part of the
conductivity appears, does not coincide with the critical value of
$\mu$ where the condensate forms. The gap forms at a \emph{larger}
value of $\mu$, and becomes more pronounced as $\mu$ increases.

\begin{figure}[t]
\begin{center}
\includegraphics[width=.45\textwidth]{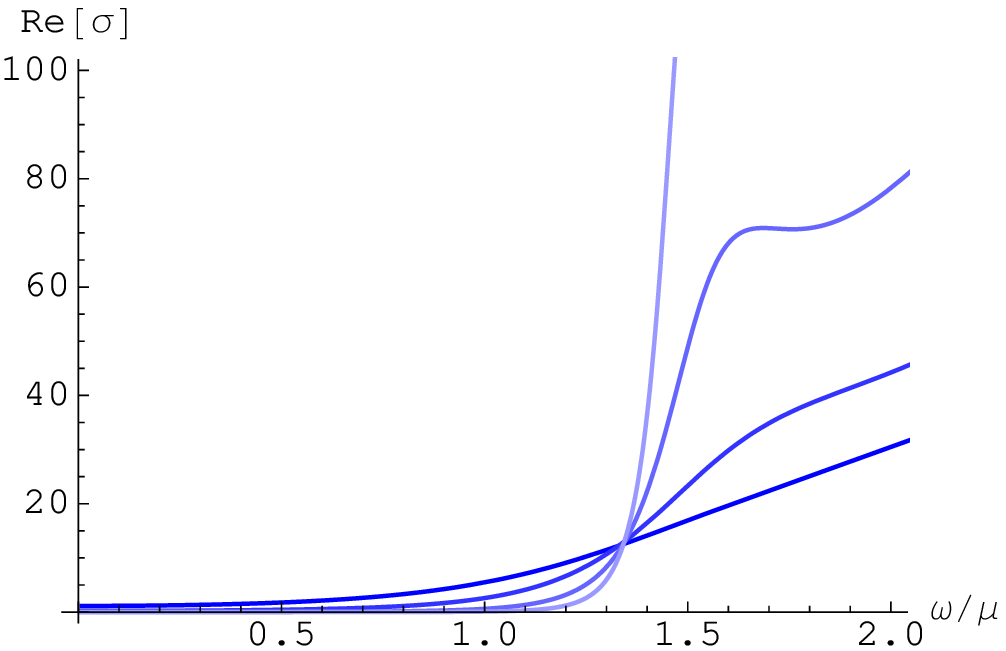}\quad
\includegraphics[width=.45\textwidth]{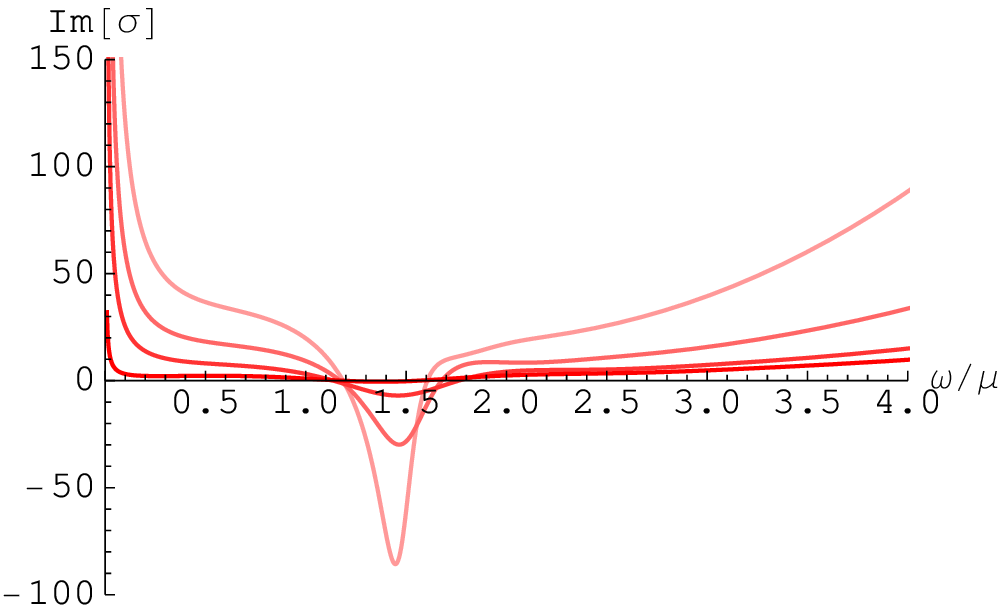}
\end{center}
\caption{The real and imaginary part of the conductivity of the D4/D8
  system, versus the dimensionless ratio~$\omega / \mu$, for various
  values of~$\mu/T$ (corresponding to the dots in
  figure~\protect\ref{f:cases}). If we define~$\omega_g$ to be the
  frequency for which the imaginary part of the conductivity has its
  minimum, this plot shows that $\omega_g \sim \mu$. The other systems
  show qualitatively similar
  behaviour.\label{f:D4D8_conductivityscaled}}
\end{figure}

In all cases we have analysed, there is a strong evidence for the
scaling of the energy gap with the critical temperature,
\begin{equation}
\frac{w_g}{T_c} \approx \text{constant}.
\end{equation}
This is most easily seen from the plots of the conductivity versus the
dimensionless ratio~$\omega/\mu$; an example is given in
figure~\ref{f:D4D8_conductivityscaled} for the D4D8 system. Following~\cite{Horowitz:2008bn}, we
define~$\omega_g$ as the frequency for which~$\Imag\sigma$ has a
minimum (following~\cite{Horowitz:2008bn}). The relation between the
critical temperature~$T_c$ and the chemical potential~$\mu$ is read
off from the plots of the condensate versus~$T/\mu$ in
figure~\ref{f:cases}: the intersection of these curves with the
horizontal axis yields the value of~$T/\mu$ for which the condensate
first appears, and hence gives~$T_c$ for a given~$\mu$.\footnote{In
  reading off the numerical coefficient, it should be noted that the
  variable along the horizontal axes of these plots is~$q T/\mu$, with
  the proportionality constant~$q$ chosen such that this reduces
  to~$1/\mu$ when~$r_T=L=1$.}  In this way we obtain
\begin{equation}
\begin{aligned}
\text{AdS}_4:\quad& T_c/\mu \approx 0.13\,, & \quad \omega_g/T_c \approx 8.4\,,\\[1ex]
\text{D2/D6}:\quad& T_c/\mu \approx 0.14\,, & \quad \omega_g/T_c \approx 8.0\,,\\[1ex]
\text{D3/D7}:\quad& T_c/\mu \approx 0.16\,, & \quad \omega_g/T_c \approx 8.1\,,\\[1ex]
\text{D4/D8}:\quad& T_c/\mu \approx 0.21\,, & \quad \omega_g/T_c \approx 6.7\,.
\end{aligned}
\end{equation}
In the conformal cases analysed thus far this ratio turned out to be
around~$8$ to within a few percent. This is in contrast to the BCS
values which are anywhere between~$3.5$ at weak coupling and $4$ in
the naive strong coupling limit (see e.g.~\cite{Marsiglio:2008a}). The
non-conformal cases give results which are higher than the BCS result
too, with the D4/D8 value deviating substantially from~$8$. Our results for
the non-abelian condensate in the $\text{AdS}_4$ black hole background
is similar to the one for the $\text{AdS}_4$ Abelian Higgs model
of~\cite{Horowitz:2008bn}.

\begin{figure}[t]
\begin{center}
\includegraphics[width=.4\textwidth]{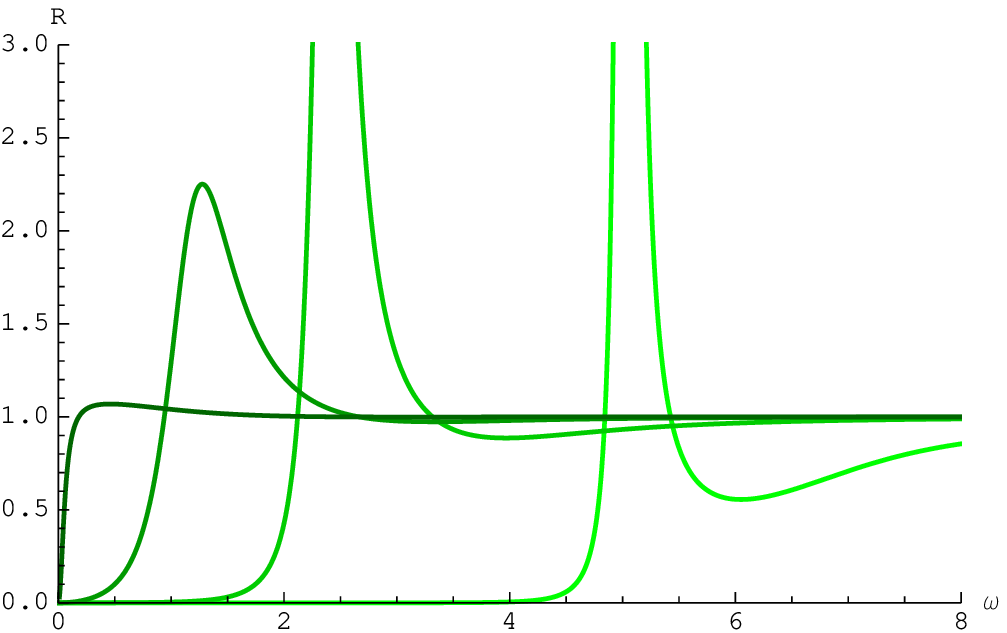}\qquad
\includegraphics[width=.4\textwidth]{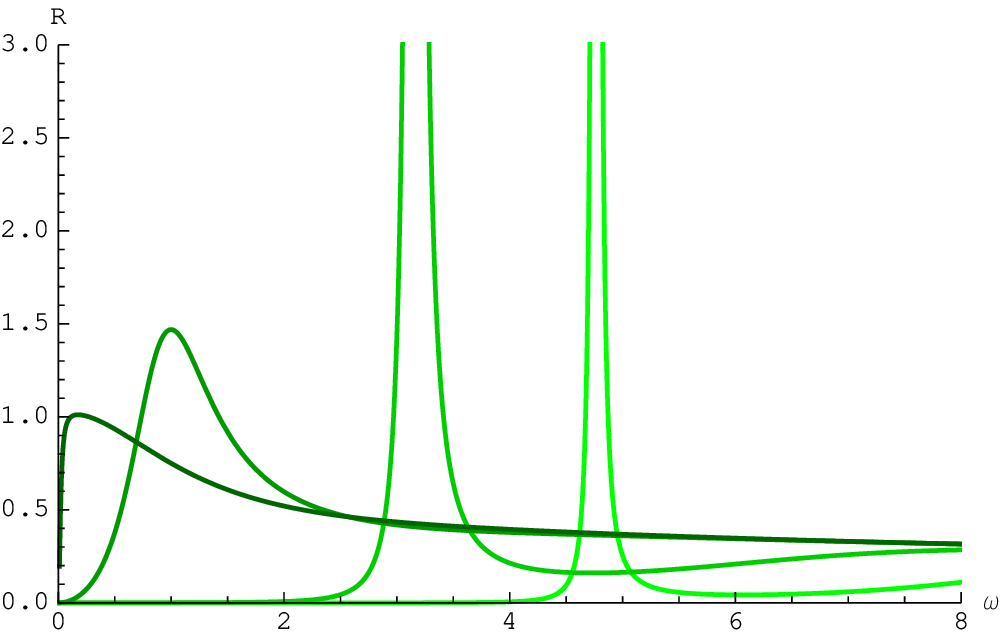}\\[2ex]
\includegraphics[width=.4\textwidth]{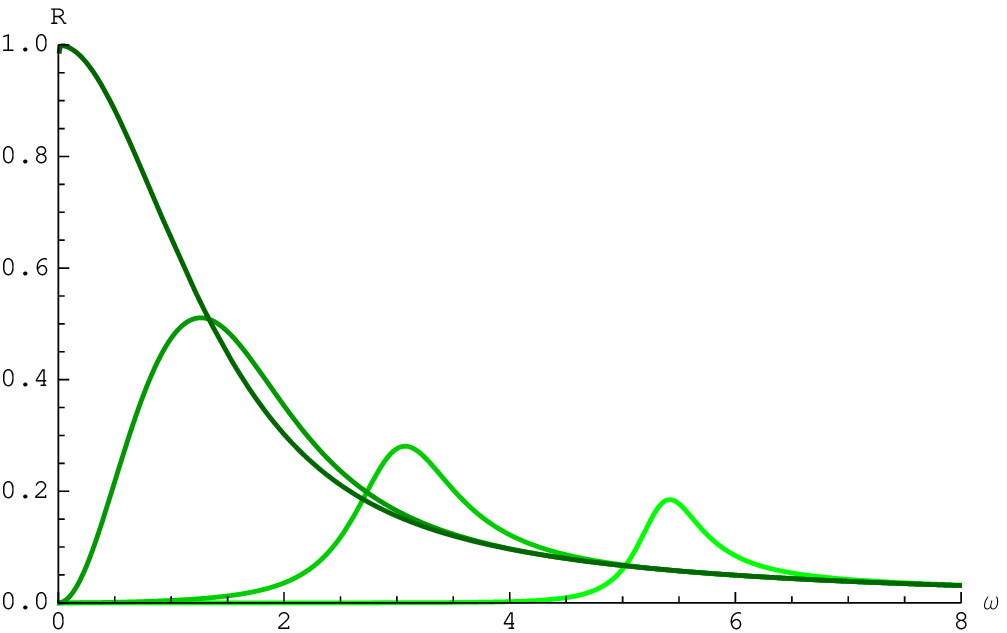}\qquad
\includegraphics[width=.4\textwidth]{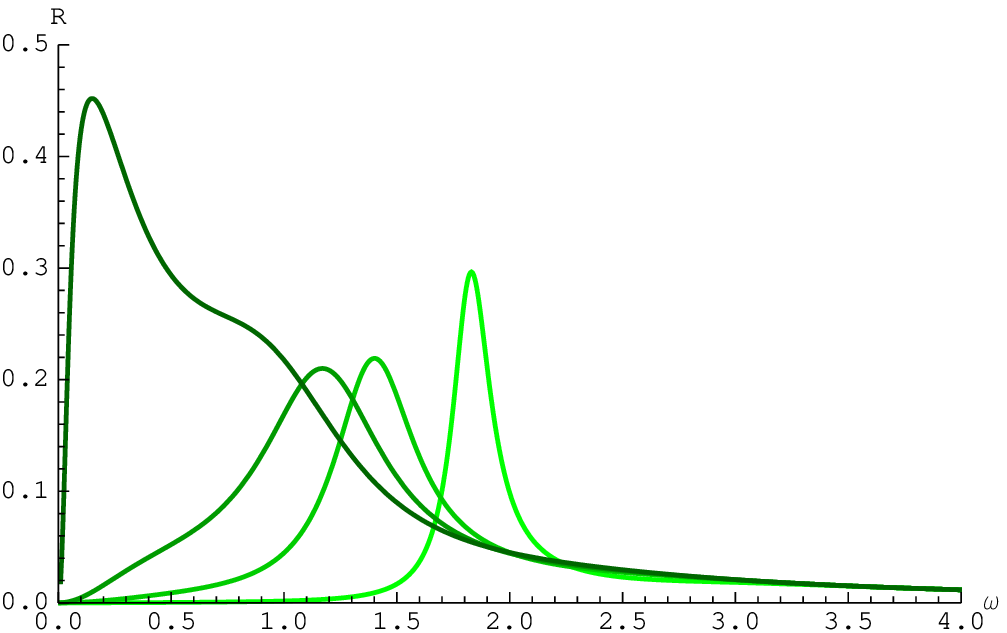}
\end{center}
\caption{Resistivity as a function of frequency (left to right, top to
  bottom: $\text{AdS}_4$, D2/D6, D3/D7 and D4/D8), for various values
  of~$\mu/T$ (corresponding to the dots in
  figure~\protect\ref{f:cases}). Lighter curves correspond to larger
  chemical potential.\label{f:resistivities}}
\end{figure}

One final way to compare the holographic superconductors with
experimental data on superconductors is to express the results in
terms of the real resistivity, defined as
\begin{equation}
R = \frac{\Real \sigma}{(\Real \sigma)^2 + (\Imag \sigma)^2}\,.
\end{equation}
Although plots of~$R$ versus~$\omega$ for the AC behaviour are not
very common in the literature, we have provided these for completeness
in figures~\ref{f:resistivities}. We will have more comments on the
behaviour of the DC resistivity in the next section.

\subsection{DC conductivity}

It has recently been noted that the $\omega=0$ limit of the real part
of the conductivity can be expressed in terms of quantities at the
horizon, rather than at infinity~\cite{Iqbal:2008by}. By expressing
the conductivity in terms of the canonical momentum, and noting that
it satisfies a trivial flow equation in the radial direction. The
conductivity can be expressed in terms of a quantity evaluated at the
horizon. A generalisation of their argument to include a dilaton reads
\begin{equation}
\Real\sigma_{\text{DC}} \propto \sqrt{\frac{-g}{g_{rr} g_{tt}}} g^{xx} e^{-\phi} \Big|_{r=r_t}\,,
\end{equation}
(where the proportionality involves an overall coupling constant
dependence which will not be important here). For the cases analysed
here, this expression yields~$\Real\sigma\propto 1$, $T^{1/3}$, $T$
and~$T^2$ respectively\footnote{The D3/D7 behaviour agrees
  with~\cite{CaronHuot:2006te} and is the same as found for the
  Abelian Higgs model in~\cite{Horowitz:2008bn}.}. It is important to
note that, except for the conformal~$\text{AdS}_4$ case, these are all
increasing functions of temperature, quite in contrast to
e.g.~ordinary metals. We should also note that while the result
of~\cite{Iqbal:2008by} was derived in the absence of any condensate,
it is also relevant for the abelian condensate, since in the
Yang-Mills approximation the fluctuation equations decouple from the
abelian condensate.

This behaviour of the DC conductivity at zero condensate is indeed
easily verified to hold true for our numerical results in
the~$\omega\rightarrow 0$ limit, thus providing a quick test of our
analysis in the abelian case. In fact, one can again use the
properties of equation~\eqref{ELfluct} explained in the appendix to
analyse the behaviour of~$\sigma$ as a function of~$T$. First, one
notes that in~$\tilde{r}$ coordinates the Green
function~\eqref{e:Green} is given by an overall power of~$TL$
multiplied with a function of~$\omega/T$. These factors of~$TL$ are
\begin{equation}
\begin{aligned}
\text{AdS}_4 &: &\quad G_R(\omega,T) &= (TL) \times \tilde{G}_R(\omega/T,\mu/T)\,,\\[1ex]
\text{D2/D6} &: &\quad G_R(\omega,T) &= (TL)^{\frac{4}{3}} \times \tilde{G}_R(\omega/T,\mu/T)\,,\\[1ex]
\text{D3/D7} &: &\quad G_R(\omega,T) &= (TL)^2 \times \tilde{G}_R(\omega/T,\mu/T)\,,\\[1ex]
\text{D4/D8} &: &\quad G_R(\omega,T) &= (TL)^3 \times \tilde{G}_R(\omega/T,\mu/T)\,.
\end{aligned}
\end{equation}
The only bare~$\omega$ dependence enters through the denominator
of~\eqref{e:conductivity_def}.

\begin{figure}[t]
\hspace{-3em}\mbox{\includegraphics[width=.37\textwidth]{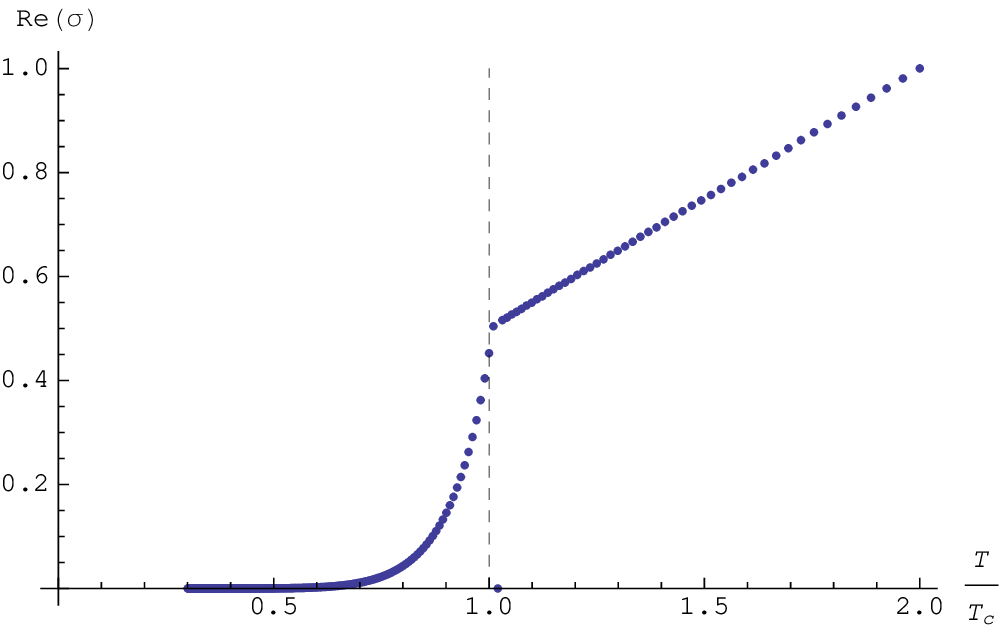}
\includegraphics[width=.37\textwidth]{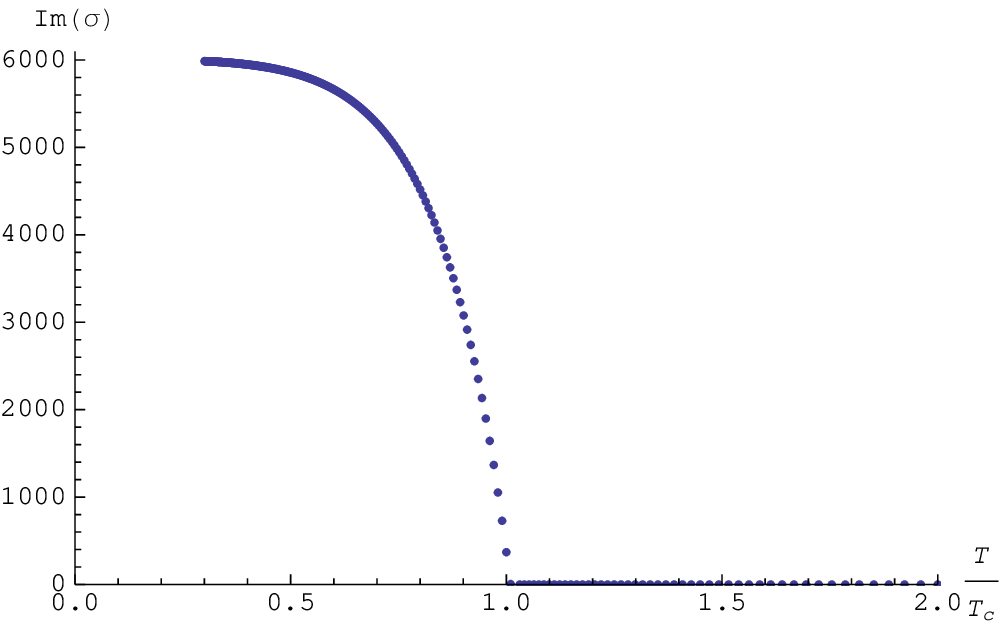}
\includegraphics[width=.37\textwidth]{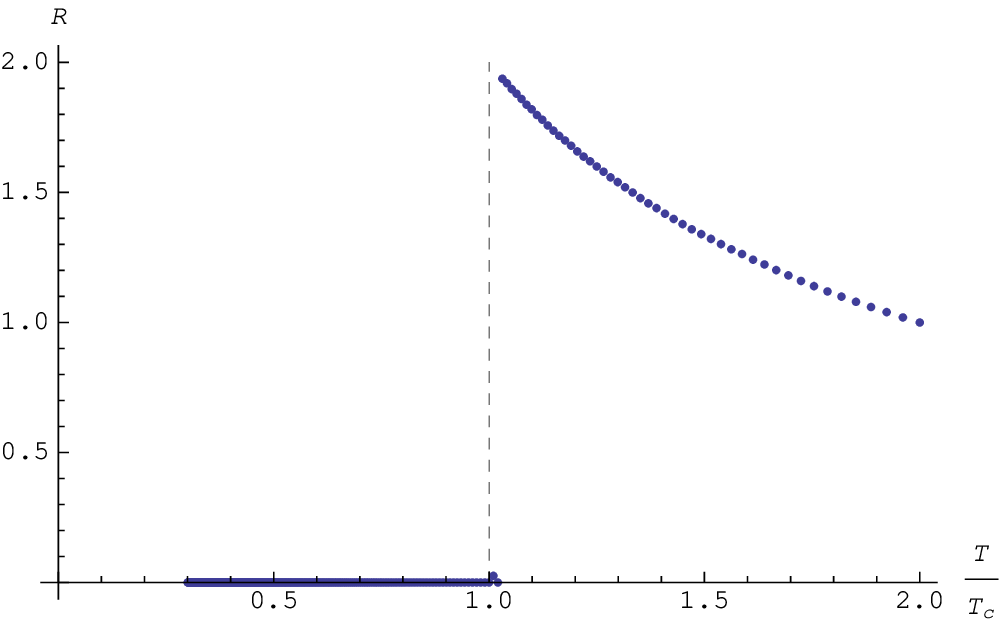}}
\caption{The real and imaginary part of the DC conductivity (or
  rather, the AC conductivity at fixed small frequency) as a function
  of temperature, and the resistivity, for the D3/D7 system. For
  temperatures $T>T_c$ the condensate becomes abelian and we recover
  the $\Real \sigma \propto T$ result
  of~\cite{CaronHuot:2006te,Iqbal:2008by}, which is the
  large-temperature limit of~\cite{Karch:2007pd}. For $T<T_c$ there is
  a regime for which $\Real \sigma$ is not yet zero, but shows a
  polynomial dependence on the temperature.\label{f:D3D7_Res_vs_T}}
\end{figure}

Once the non-abelian condensate is turned on, the analysis
of~\cite{Iqbal:2008by} can no longer be applied. The coupling between
the fluctuation and the gauge field component~$A_x$ implies that it is
now no longer true that the radial evolution of the canonical momentum
is trivial. Hence it is no longer possible to express the conductivity
in terms of a quantity evaluated at the horizon. Nevertheless, it is
possible to analyse the temperature dependence of the conductivity in
the $\omega\rightarrow 0$ limit numerically. The scaling properties
imply that~$\mu/T \propto \omega/T$. Figure~\ref{f:D3D7_Res_vs_T}
displays the behaviour of the conductivity
and resistivity in this non-abelian phase.\footnote{The metallic~$\Real\sigma
  \propto 1/T^2$ behaviour at small temperatures, found
  in~\cite{Karch:2007pd}, is invisible in the Yang-Mills
  truncation. It is, however, unclear whether it will show up before
  the transition to the non-abelian condensate takes place.}

Generically, we see that superconductors of the holographic type
exhibit a behaviour of the resistivity which is similar to that of
semi-conductors, i.e.~$R$ decreases with temperature above the
critical temperature. A metallic behaviour has so far only been found
for small temperatures in the model of~\cite{Karch:2007pd}.

\subsection{Comments on connections with real-world superconductors}
\label{s:realworld_em}

\begin{figure}[t]
\begin{center}
\vspace{1ex}
\includegraphics[width=.45\textwidth]{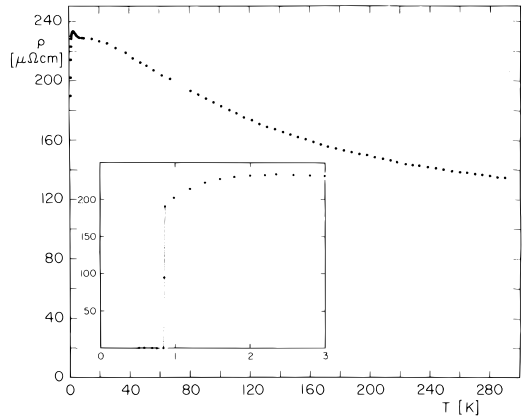}
\vspace{-2ex}
\end{center}
\caption{Resistivity of the heavy fermion compound
  $\text{UBe}_{13}$. Figure taken from~\cite{Ott:1983a}.
\label{f:HFresistivity}}
\end{figure}

We have already seen in section~\ref{s:realworld_cv} that weakly
coupled and strongly coupled superconductors exhibit very different
behaviour for their specific heat. As far as the electric resistivity
of these materials is concerned, while all of them have zero
resistivity for $T<T_c$, when $T>T_c$, their behaviour again differs
significantly. The weakly coupled superconductors mainly behave as
``metals'' above $T_c$ (i.e.~with a resistivity that grows with~$T$ as
a polynomial of $T$). On the other hand, just above $T_c$ strongly
coupled superconductors behave either as metals or as semiconductors, with a
resistivity that decreases with increasing temperature (see
figure~\ref{f:HFresistivity}).

Just like real superconductors, all holographic models exhibit zero
conductivity when $T< T_c$.  Interestingly, the temperature at which
the resistivity becomes strictly zero is lower than the temperature
$T_{c}$ at which the non-abelian condensate becomes the true ground
state. This is most clearly visible in
figure~\ref{f:D3D7_Res_vs_T}. The small temperature range $T_0 < T <
T_{c}$ corresponds to a transition region where the resistivity very
quickly drops from the normal-phase value to zero (the drop is more
clearly visible in the real part of the conductivity, as the
resistivity is suppressed by the large imaginary part).

When $T>T_{c}$ only $A_0\neq 0$, and hence only the abelian
condensate~$\rho$ determines the behaviour of the system. The
Yang-Mills truncation leads to a conductivity which is
\emph{independent} of the condensate (or chemical potential), due to
the fact that the electromagnetic fluctuations fully decouple from the
condensate in the abelian approximation.  In this approximation,
the system behaves as a semi-conductor, with a resistivity which decreases
with increasing temperature (see figure~\ref{f:D3D7_Res_vs_T}).  The
DBI approximation, on the other hand, couples the fluctuations to the
condensate, and leads to a condensate-dependent result for the
conductivity~\cite{Karch:2007pd}, 
\begin{equation}
\sigma^2 = a T^2 + \frac{\rho^2}{\lambda}
\frac{b}{T^4} \,,
\end{equation}
where~$a$ and~$b$ are numerical constants and~$\lambda$ is the
't~Hooft coupling.  The first, condensate-independent term originates
from charge carriers that are thermally produced in charge neutral
pairs, while the second term originates from the charge carriers
introduced by the abelian condensate. The first term was visible in the
Yang-Mills approximation and leads to semiconducting behaviour (an
increase of temperature enhances thermal pair production), while the
second term describes metallic behaviour.


The system thus behaves as a metal below the temperature $T<T_{*}$,
\begin{equation}
T_{*} = \left(\frac{2\rho^2 b}{a \lambda}\right)^{1/6}\,,
\end{equation}
which is governed by $\alpha'$ corrections to the Yang-Mills
action. As the system is cooled, the question is whether metallic
behaviour will kick in before the superconducting phase or
not. Obviously, this depends on whether $T_{c}$ is smaller or larger
than $T_{*}$.  The requirement that $T_* > T_c$, i.e.~that the system
behaves as a metal above~$T_c$, will put constraints on $\lambda$, and
determine if it is compatible with the requirement that $\lambda \ll 1$.
Of course, one should keep in mind that our analysis for $T<T_c$ has
been done in the (non-abelian) Yang-Mills approximation, while the
analysis for $T>T_c$ was done in the DBI approximation. Therefore, the
first step one should take is to see how higher-derivative corrections
could change the critical temperature. We leave this issue for future
investigation.

\section*{Acknowledgements}

We would like to thank Imperial College for hospitality while parts of
this paper were being written.

\appendix
\section{Appendix: computing temperature dependence}
\label{s:scaling}

While the coupled system of equations~\eqref{e:nonlinear} is
complicated and does not generically admit analytic solutions, it does
have a useful scaling symmetry (both in the conformal and nonconformal
cases), which simplifies our analysis.  When considering a
non-conformal systems we deal with three dimensionful parameters, the
scale $L$, the temperature~$T$ and the chemical potential~$\mu$. One
might thus expect that the dimensionless condensate is a function of
the two dimensionless combinations $TL$ and $\mu L$.  However, the
equations~\eqref{e:nonlinear} allow us to reduce this dependence to
only one dimensionless combination $\mu/T$, with all $TL$ dependence
following from general arguments.  This is a consequence of a scaling
symmetry of~\eqref{e:nonlinear}, even when the
metric~\eqref{e:bgmetric} does not have conformal symmetry.

To see this, let us introduce a dimensionless radial
coordinate~$\tilde{r} = r / r_T$. In this coordinate the equations
that determine the condensate read
\begin{equation}
\label{e:dimlesseqs}
\begin{aligned}
\partial_{\tilde{r}} \Big( h_1(\tilde{r})\, \partial_{\tilde r}
A_0^{(3)}\Big) &= 4 (A_3^{(1)})^2 A_0^{(3)} h_2(\tilde{r})\,
T^{-2}\,,\\[1ex]
\partial_{\tilde{r}} \Big( h_3(\tilde{r})\, \partial_{\tilde r}
A_3^{(1)}\Big) &= 4 (A_0^{(3)})^2 A_3^{(1)} h_4(\tilde{r})\,
T^{-2}\,,
\end{aligned}
\end{equation}
where the~$h_i(\tilde{r})$ are functions of~$\tilde{r}$ only.  That
is, all $T$-dependence sits in the explicit~$T^{-2}$ factor of the
terms on the right-hand side, and there is no bare dependence on~$L$
anymore. Solutions to the equations now take the form
\begin{equation}
\label{e:A0expdimless}
A_0^{(3)} = \mu + \rho\, r^\alpha + \ldots
= T \Big( \tilde{\mu} + \tilde{\rho} \tilde{r}^\alpha + \ldots
\Big) =: T \tilde{A}_0^{(3)}\,,
\end{equation}
and similar for~$A_3^{(1)}$ (see e.g.~\eqref{e:A0dimlessD4D8} for the
D4/D8 case). Here $\alpha$ is a model-dependent constant and we have
defined
\begin{equation}
\label{e:defmurhotilde}
\tilde{\mu} := \frac{\mu}{T} \quad\text{and}\quad
\tilde{\rho} := \frac{\rho\, (r_T)^\alpha}{T}\,.
\end{equation}
From the fact that the equations~\eqref{e:dimlesseqs} are independent
of~$T$ and $L$ when expressed in terms of the $\tilde{A}_0^{(3)}$ and
$\tilde{A}_3^{(1)}$ variables, we can then conclude that the
dimensionless quantity $\tilde{\rho}$ only depends on $\tilde{\mu}$,
not on the dimensionless combination $TL$. For the D4/D8 system, for
instance, the expansion reads
\begin{equation}
\label{e:A0dimlessD4D8}
\text{D4/D8}:\quad A_0^{(3)} = T \left[ \frac{\mu}{T} - \frac{\rho}{T^{5/2} (TL)^{3/2}}
  \tilde{r}^{3/2} + \ldots\right] =: T \Big[ \tilde{\mu} - \tilde{\rho}
\,\tilde{r}^{3/2} + \ldots\Big]\,.
\end{equation}

So we see that even when the theory is not conformally
invariant\footnote{In the conformal D3/D7 case, this story simplifies
  and the dimensionless condensate~$\rho/T$ is independent of $TL$,
  but the argument presented here shows that the story is somewhat
  more complicated in the non-conformal cases, where the naive
  dimensionless combination~$\rho/T^{(\alpha+1)}$ still depends on
  $TL$, albeit only in the simple way dictated by~\eqref{e:defmurhotilde}.}, the scaling symmetry
of the equations allows us to determine the dependence of the
dimensionless condensate on~$TL$. We also note that if one were to
include higher derivative terms into the equations of motion (which
for example follow from the DBI action), then the above symmetry would
not necessarily continue to be present. In practice, we have used the
scaling symmetry to verify numerical stability of the condensate
solutions: plots of~$\tilde{\mu}$ against~$\tilde{\rho}$ should be
independent of~$TL$.

The scaling argument also allows us to determine e.g.~the temperature
dependence of the action by changing~$\mu$ instead of~$r_T$. The idea
here is to compute the action in~$\tilde{r}$ coordinates at a fixed
value~$r_T=1$. This will yield~$S(\mu,T) = r_T^\beta
\tilde{S}(\mu/T, r_T=1)$ for some~$\beta$, with the overall power of~$r_T$
simply arising from the coordinate transformation~$r = r_T
\tilde{r}$. We will see this at work in section~\ref{s:free_energy}.

\vfill\eject

\begingroup\raggedright\endgroup

\end{document}